\documentclass{aa}

\usepackage{amssymb,amsmath,hyperref,txfonts,amsfonts}
\usepackage{xcolor}
\usepackage{mathtools}
\usepackage{mathrsfs}
\usepackage{ulem} 

\hypersetup{colorlinks,citecolor=blue,linkcolor=blue,urlcolor=blue}

\newcommand\sbullet[1][.5]{\mathbin{\vcenter{\hbox{\scalebox{#1}{$\bullet$}}}}}

\def\sgra{\object{Sgr~A$^{\ast}$}\xspace}
\def\m87{\object{M87$^{\ast}$}\xspace}

\newcommand\eht{{EHT}\xspace}
\newcommand\chandra{\textsl{Chandra}\xspace}

\defcitealias{Issaoun2022}{Issaoun, Wielgus et al. 2022}

\begin{document}

   \title{Orbital motion near Sagittarius A$^*$}

   \subtitle{Constraints from polarimetric ALMA observations}

   \author{M. Wielgus
          \inst{1,2,3},
          M. Moscibrodzka\inst{4},
          J. Vos\inst{4},
          Z. Gelles\inst{5,3},
          I. Martí-Vidal\inst{6,7},
          J. Farah\inst{8,9},
          N. Marchili\inst{10,1},\\
          C. Goddi\inst{11,12},
          \and
          H.~Messias\inst{13}
          }

   \institute{Max-Planck-Institut f\"ur Radioastronomie, Auf dem H\"ugel 69, D-53121 Bonn, Germany\\
         \email{maciek.wielgus@gmail.com}
         \and
         Nicolaus Copernicus Astronomical Centre, Polish Academy of Sciences, Bartycka 18, 00-716 Warsaw, Poland
         \and
             Black Hole Initiative at Harvard University, 20 Garden Street, Cambridge, MA 02138, USA
             \and
             Department of Astrophysics/IMAPP, Radboud University,P.O. Box
             9010, 6500 GL Nijmegen, The Netherlands
             \and
             Center for Astrophysics | Harvard \& Smithsonian, 60 Garden Street, Cambridge, MA 02138, USA
             \and
             Departament d'Astronomia i Astrof\'{\i}sica, Universitat de Val\`encia, C. Dr. Moliner 50, E-46100 Burjassot, Val\`encia, Spain
             \and
             Observatori Astronòmic, Universitat de Val\`encia, C. Catedr\'atico Jos\'e Beltr\'an 2, E-46980 Paterna, Val\`encia, Spain
             \and
             Las Cumbres Observatory, 6740 Cortona Drive, Suite 102, Goleta, CA 93117-5575, USA
             \and
Department of Physics, University of California, Santa Barbara, CA 93106-9530, USA
\and
             Italian ALMA Regional Centre, INAF-Istituto di Radioastronomia, Via P. Gobetti 101, I-40129 Bologna, Italy
             \and
            Dipartimento di Fisica, Università degli Studi di Cagliari, SP Monserrato-Sestu km 0.7, I-09042 Monserrato, Italy
            \and
            Universidade de São Paulo, Instituto de Astronomia, Geofísica e Ciências Atmosféricas, Departamento de Astronomia, São Paulo, SP 05508-090, Brazil
            \and
           Joint ALMA Observatory, Alonso de Cordova 3107, Vitacura 763-0355, Santiago de Chile, Chile\\ 
             }

\abstract
 {We report on the polarized light curves of the Galactic Center supermassive black hole Sagittarius A$^*$, obtained at millimeter wavelength with the Atacama Large Millimeter/submillimeter Array (ALMA). The observations took place as a part of the Event Horizon Telescope campaign. We compare the observations taken during the low variability source state on 2017 Apr 6 and 7 with those taken immediately after the X-ray flare on 2017 Apr 11. For the latter case, we observe rotation of the electric vector position angle with a~timescale of $\sim\,70$\,min. We interpret this rotation as a~signature of the equatorial clockwise orbital motion of a~hot spot embedded in a~magnetic field dominated by a dynamically important vertical component, observed at a low inclination $\sim\,20^\circ$. The hot spot radiates strongly polarized synchrotron emission, briefly dominating the linear polarization measured by ALMA in the unresolved source. Our simple emission model captures the overall features of the polarized light curves remarkably well. Assuming a~Keplerian orbit, we find the hot spot orbital radius to be $\sim$\,5 Schwarzschild radii. We observe hints of a positive black hole spin, that is, a prograde hot spot motion. Accounting for the rapidly varying rotation measure, we estimate the projected on-sky axis of the angular momentum of the hot spot to be $\sim\,60^\circ$ east of north, with a 180$^\circ$ ambiguity. These results suggest that the accretion structure in \sgra is a~magnetically arrested disk rotating clockwise.  }

   \keywords{black holes -- galaxies: individual: Sgr A* -- Galaxy: center -- techniques: interferometric }
   
   \titlerunning{Orbital motion near Sgr A$^*$}
   \authorrunning{Wielgus et al.}

\maketitle

\section{Introduction}

Sagittarius A$^\ast$ (\sgra) is associated with a supermassive black hole located in the center of our galaxy, with a mass $M_{\sbullet[1.35]} \approx 4\times$10$^6\,M_{\odot}$ \citep{Do2019,Gravity2022,SgraP1}. The source exhibits rapid variability across the electromagnetic spectrum, with particularly strong flaring events in the infrared \citep[IR; e.g.,][]{Genzel2003,Eckart2006,Do2019_IRflare} and X-ray \citep[e.g.,][]{Baganoff2001,Porquet2003,Haggard2019}, during which the observed flux density rises by 1-2 orders of magnitude. While the detailed physics of these energetic episodes is not fully understood, magnetic reconnection in a radiatively inefficient magnetized accretion flow \citep{Yuan2014} constitutes a plausible theoretical framework to interpret the flaring activity of \sgra \citep{Yuan2003,Dodds2010,Dexter2020}. In a reconnection event, the magnetic energy can be rapidly dissipated, heating plasma locally. The exhaust of the reconnection can form a transient feature in the orbiting accretion flow -- a hot spot of low-density plasma confined in a flux tube of a vertical magnetic field \citep[e.g.,][]{Porth2021,Ripperda2022}. It has been proposed that investigating the dynamics of such transient flow features would constitute a powerful probe into the physics of gravity and accretion \citep{Broderick2005,Broderick2006, Hamaus2009, Zamaninasab2010}.

\begin{figure*}
    \centering
    \includegraphics[width=0.950\textwidth,trim=-0.0cm 0.0cm 0cm 0cm,clip]{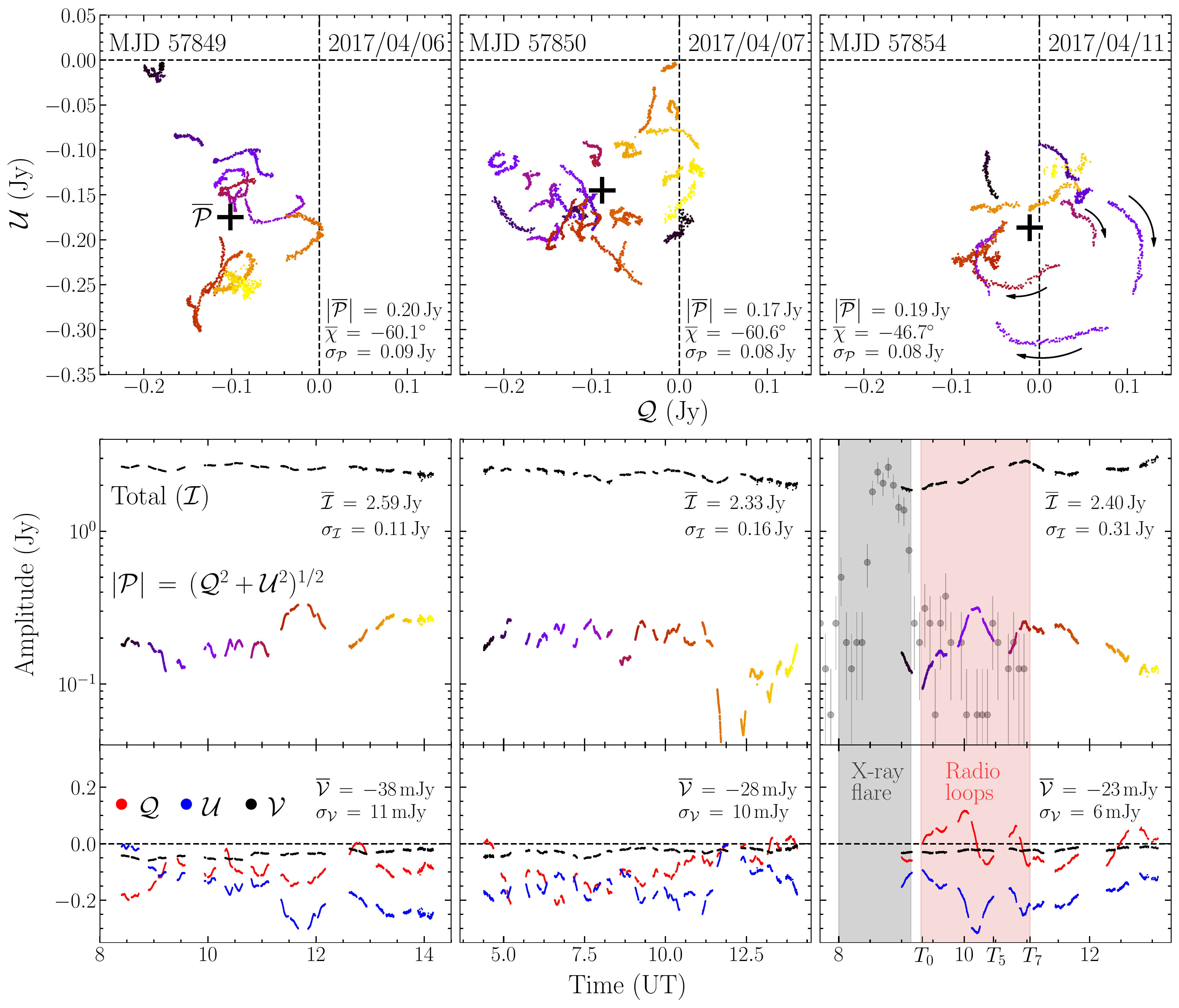}
    \caption{Full Stokes 229\,GHz ALMA light curves of \sgra obtained on 2017 Apr 6, 7, and 11. \textit{Top row:} linear polarization represented on a $\mathcal{Q}$-$\mathcal{U}$ plane. The color of the curve indicates the time from the beginning of the observation. Crosses "+" show the locations of the daily mean linear polarization $\overline{P}$ with the EVPA $\overline{\chi}$. The pattern observed on 2017 Apr 11 suggests the presence of coherent clockwise $\mathcal{Q}$-$\mathcal{U}$ loops, emphasized with arrows. \textit{Bottom rows:} full Stokes \sgra light curves. On Apr 11, ALMA started observing immediately after the X-ray flare (gray band region). \chandra X-ray data are plotted in counts per second and were arbitrarily scaled. The red-shaded region corresponds to the period in which polarimetric loops are apparent.}
    \label{fig:general_LC}
\end{figure*}

Studying hot spots associated with \sgra flares became possible with the GRAVITY near-IR interferometer, achieving an astrometric precision of tens of microarcseconds \citep{Gravity2017}. Several observations of hot spots' signatures during the \sgra flaring state were reported \citep{gravity_loops_2018}. In all cases, a clockwise, nearly circular motion of the brightness centroid on a timescale of 0.5-1.0\,h was detected, which is consistent with the interpretation of an orbiting hot spot in an innermost region of the accretion flow, viewed at low inclination. Simultaneously with the brightness centroid rotation, polarimetric signatures of a hot spot threaded by the magnetic field were found in the form of an electric vector position angle (EVPA) rotation \citep[][]{gravity_loops_2018, Baubock2020,Jimenez2020}. 

The activity of \sgra appears less dramatic at millimeter wavelengths, with a flux density increase of several tens of percent lagging behind the IR and X-ray flares \citep{Marrone2008,wielgus22}. Resolving relative motions with angular resolution comparable to that of GRAVITY is only possible at millimeter wavelengths using very-long-baseline interferometry (VLBI) with the Event Horizon Telescope (EHT) array \citep{paperii}. However, the EVPA rotation in the unresolved compact source can be traced with connected-element interferometric arrays such as ALMA or the SubMillimeter Array (SMA). Hints of the coherent EVPA evolution in the SMA observations were already reported by \citet{Marrone2006JPhCS}.

In this paper we present full Stokes light curves of millimeter emission from \sgra observed by ALMA on 2017 Apr 6, 7, and 11. We focus on the Apr 11 observations, in the period immediately following an X-ray flare observed by \chandra \citep{SgraP2,wielgus22}. We identify polarimetric signatures of an orbiting hot spot and, using a simple emission model, we infer system parameters broadly consistent with the findings of GRAVITY  \citep{gravity_loops_2018}.


\section{Observations and data properties}
\label{sec:obsdata}
\subsection{ALMA observations}
\sgra was observed with ALMA as a part of the EHT campaign in 2017. ALMA participated in the EHT VLBI observations as a phased array \citep{Goddi2019,SgraP1}. In parallel to the VLBI data reduction pipeline, the phased array observations were processed to recover the connected interferometer ALMA-only measurements \citep{Goddi2021}. The initial calibration was performed following the ALMA QA2 procedures described in \cite{Goddi2019}, benefiting from the long duration of observing tracks and the utilization of multiple calibrator sources that are necessary to enable full Stokes VLBI imaging with the EHT \citepalias[\citealt{martividal2016,Paper7};][]{Issaoun2022}. 
Subsequently, an additional correction of time-dependent amplitude gains was performed, after modeling the time variability of the compact \sgra source against the static parsec-scale Galactic
Center minispiral \citep{Lo1983, Mus2022}. This self-calibration procedure corresponds to the A1 data reduction pipeline described in \citet{wielgus22}, where a detailed discussion of the related algorithms, extraction of the compact unresolved source signal, and quality control can be found.
The resulting full Stokes light curves of \sgra have a~cadence of 4\,s and an exquisite signal-to-noise ratio (S/N) formally reaching $\sim\,1000$ (for the discussion of systematic uncertainties, see Appendix E in \citealt{Goddi2021}, and Sections 4-5 in \citealt{wielgus22}). The light curves correspond to four frequency bands of 2\,GHz width each, with central frequencies of 213.1, 215.1, 227.1, and 229.1\,GHz.

\subsection{Mean light curves properties}
\label{sec:mean_lc_properties}

The light curves corresponding to all three observing epochs (2017 Apr 6, 7, and 11) are shown in Fig.~\ref{fig:general_LC}. Time-averaged parameters of the Stokes components are also provided. The observations took place during a period of a particularly low total intensity of \sgra. The properties of the total intensity (Stokes $\mathcal{I}$) light curves are discussed in detail in \citet{wielgus22}. Here, we focus on the linear polarization (LP), $\mathcal{P} = \mathcal{Q} + i\,\mathcal{U}$, and circular polarization (CP), $\mathcal{V}$. There is a good consistency between the three observing days on the average LP, corresponding to 7-8\% of the total intensity, as well as the average CP, corresponding to about -1\% of the total intensity. The EVPA (denoted as $\chi$), defined as $ \chi = 0.5 \arg\,\mathcal{P}$,
corresponds to $-57\pm11$\,deg, with Apr 6-7 corresponding to the average of $\chi \approx -60^\circ$, and Apr 11 corresponding to the average $\chi \approx-47^\circ$ in the 229.1\,GHz band. The EVPA differs between frequency bands because of the significant rotation measure (RM), see Appendix \ref{app:RM}.
The measured polarimetric parameters are broadly consistent with the historical measurements, for example from \citet{Bower2003,Marrone2008,Munoz2012,Bower2018}. 
\subsection{Polarimetric loops}
\label{sec:loops_intro}

\begin{figure}[t]
    \centering
    \includegraphics[width=0.95\columnwidth]{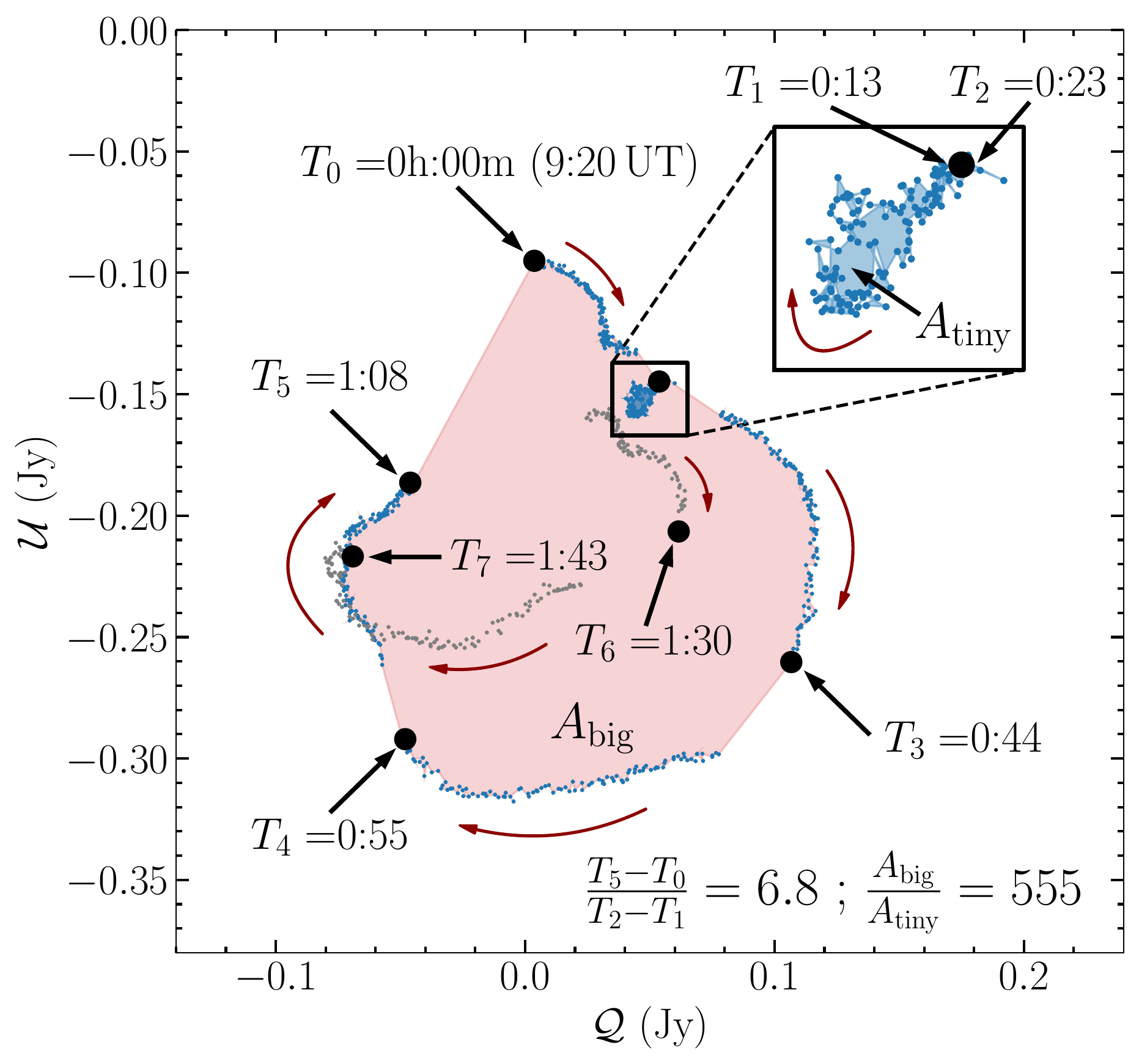}
    \caption{Polarimetric loops observed by ALMA at 229\,GHz on 2017 Apr 11. The observations began at $T_0 = $\,9:20\,UT (MJD 57854.39), 30\,min after the peak of the X-ray flare. We highlight the small inner loop $A_{\rm tiny}$. Following the full loop $A_{\rm big}$ between $T_0$ and $T_5$, a similar pattern continues between $T_5$ and $T_7$, with a decreased period and reduced LP.
}
    \label{fig:loops_data}
\end{figure}

A notable difference between the LP signatures on 2017 Apr 6-7 and 2017 Apr 11 is the presence of coherent loops on the $\mathcal{Q}$-$\mathcal{U}$ plane in the latter data set -- see the top row of Fig.~\ref{fig:general_LC}. These signatures appear in the time period following the X-ray flare detected by \chandra, which peaked at 8:48\,UT. After about 2\,h this "loopy period" ends, and the variability of the LP component returns to a more stochastic character, similar to that observed on 2017 Apr 6-7. This suggests a causal connection between the X-ray flare and the $\mathcal{Q}$-$\mathcal{U}$ plane loops observed at a millimeter wavelength. In Fig.~\ref{fig:loops_data} we inspect data corresponding to the loopy period. We observe two large clockwise loops, first $A_{\rm big}$ with a~period slightly above 1\,h, executed between $T_0 = $\,9:20\,UT and $T_5 = T_0 + 68$\,min, and a second one with a reduced period, observed between $T_5$ and $T_7 = T_5 + 35$\,min. The first loop is very close to a round shape, with asymmetry, defined as the largest ratio of perpendicularly projected diameters, $A \sim\!1.2$. The second loop is smaller and significantly more distorted. We also observe a small inner loop $A_{\rm tiny}$, traversed in just $\sim$10\,min between $T_1$ and $T_2$. 
The ratio of the first large loop area to the small loop area is $R_a \sim\!550$, and the ratio of the full loop period to the time spent inside a small loop is $R_t \sim\!7$. We interpret these loops as signatures of an orbiting hot spot and use the three dimensionless observables ($A$, $R_t$, and $R_a$) to crudely constrain the system geometry -- see Appendix \ref{app:semi-analytic}.

\section{Interpretation and modeling}
\label{sec:modeling}

\subsection{General model specification}
\label{sec:general_model}

ALMA light curves from the 2017 EHT campaign correspond to the observed Stokes components averaged over the region unresolved by the instrument $\sim$1\,arcsec, or $\sim\!10^5r_{\rm g}$ for $r_{\rm g} = GM_{\sbullet[1.35]}/c^2$. The millimeter flux density is strongly dominated by the most compact region of the accretion flow, confined to the field of view of $\sim$100\,$\mu$as, or $\sim$20 $r_{\rm g}$  \citep{doeleman2008,SgraP2}.
The total intensity image of \sgra in a nonflaring state corresponds to a ring of $\sim\!50\,\mu$as diameter \citep{SgraP3,SgraP4}, which is consistent with the expected appearance of a Kerr black hole surrounded by a radiatively inefficient accretion flow \citep{SgraP5}. We refer to this feature as the observable shadow of a black hole \citep{SgraP6}. Its total millimeter flux density fluctuates as a red noise stochastic process, which is qualitatively (but not necessarily quantitatively, see \citealt{SgraP5} for details) consistent with the variability of the numerical models of turbulent accretion flow \citep{Georgiev2022,wielgus22}. We assume that this single stochastic component, fluctuating around $\mathcal{I}_{\rm shadow}\!\equiv\!\mathcal{I}_{\rm sha} \!\approx\!2.4$\,Jy, $|\mathcal{P}_{\rm shadow}|\!\equiv\!|\mathcal{P}_{\rm sha}|\!\approx\!0.2$\,Jy and $\mathcal{V}_{\rm shadow}\!\equiv\!\mathcal{V}_{\rm sha}\!\approx\!-0.03$\,Jy suffices to understand light curves observed outside of the loopy period. 

To interpret observations in the loopy phase, we consider a presence of an additional component, an equatorial orbiting hot spot, very similar to that envisioned by \citet{Broderick2005}. From the size of the $\mathcal{Q}$-$\mathcal{U}$ loop shown in Fig.~\ref{fig:loops_data}, we conclude a polarized flux density of $|\mathcal{P}_{\rm hot\,spot}| \equiv |\mathcal{P}_{\rm hsp}| \lesssim 0.15$\,Jy, and $\mathcal{I}_{\rm hot\,spot} \equiv \mathcal{I}_{\rm hsp}  \lesssim\!0.3$\,Jy, assuming a strong fractional polarization of 50\%. This implies that while $\mathcal{I}_{\rm sha} \gg \mathcal{I}_{\rm hsp}$, we may expect $|\mathcal{P}_{\rm hsp}| \approx |\mathcal{P}_{\rm sha}|$. Hence, our best chance of characterizing the hot spot component is through LP, rather than through the total intensity. Furthermore, we assume that during the 68\,min between $T_0$ and $T_5$ in Fig.~\ref{fig:loops_data} we can approximate $\mathcal{P}_{\rm sha}$ with a constant value with $|\mathcal{P}_{\rm sha}| \approx 0.2$\,Jy, and since the measured LP is a coherent sum of the polarization of the two components, we can assume that a variation in the observed LP in the loopy period is strongly dominated by the rapidly changing hot spot component, $\mathcal{P}_{\rm hsp}$. We note that the loop component tends to average out when coherently averaged in time, explaining the roughly consistent daily mean polarization observed by ALMA on 2017 Apr 6, 7, and 11.

From the physical point of view, a larger relative importance of the hot spot component in LP rather than in total intensity seems very reasonable. We expect the energized hot spot to indicate an increased temperature, lower density, and lower optical depth than the surrounding flow, which should result in an increase in the fractional polarization of the produced radiation \citep{Rybicki1979}.
Furthermore, we generally expect the hot spot to be smaller than the observable shadow \citep[e.g.,][]{Baubock2020}, and hence to experience less depolarization when coherently averaged over a spatially varying magnetic field structure. 

As a hot spot moves along its orbit, the trajectories of photons reaching a distant observer are lensed because of the spacetime curvature, while the transported flux density is affected by the Doppler effect and gravitational redshift. The polarization vector, perpendicular to the direction of propagation and the magnetic field in the emission frame, is parallel-transported along the null geodesics and influenced by Faraday effects (e.g., see Appendix \ref{app:RM}). Hence, the imprint of the system geometry (spacetime curvature, observer's inclination, and magnetic field configuration) on the polarimetric pattern on a $\mathcal{Q}$-$\mathcal{U}$ plane can be used to constrain parameters of the source model. Under the assumption of axisymmetry, and particularly for low inclinations, these signatures resemble circular loop patterns. A pedagogical discussion of the observable LP signatures and impact of different effects (Doppler and lensing) and parameters (magnetic field topology and emitter velocity) is given in Section 3 of \citet{Narayan2021}, and more detailed analyses can be found in \citet{Gelles2021} and in \citet{Vos2022}. We notice that the reported ALMA $\mathcal{Q}$-$\mathcal{U}$ loop is traversed clockwise, consistent with the observations of \citet{gravity_loops_2018}. In all models considered in this paper, this implies a clockwise physical motion of the orbiting hot spot in the sky.

\begin{figure*}[t]
    \centering
    \includegraphics[width=0.95\linewidth]{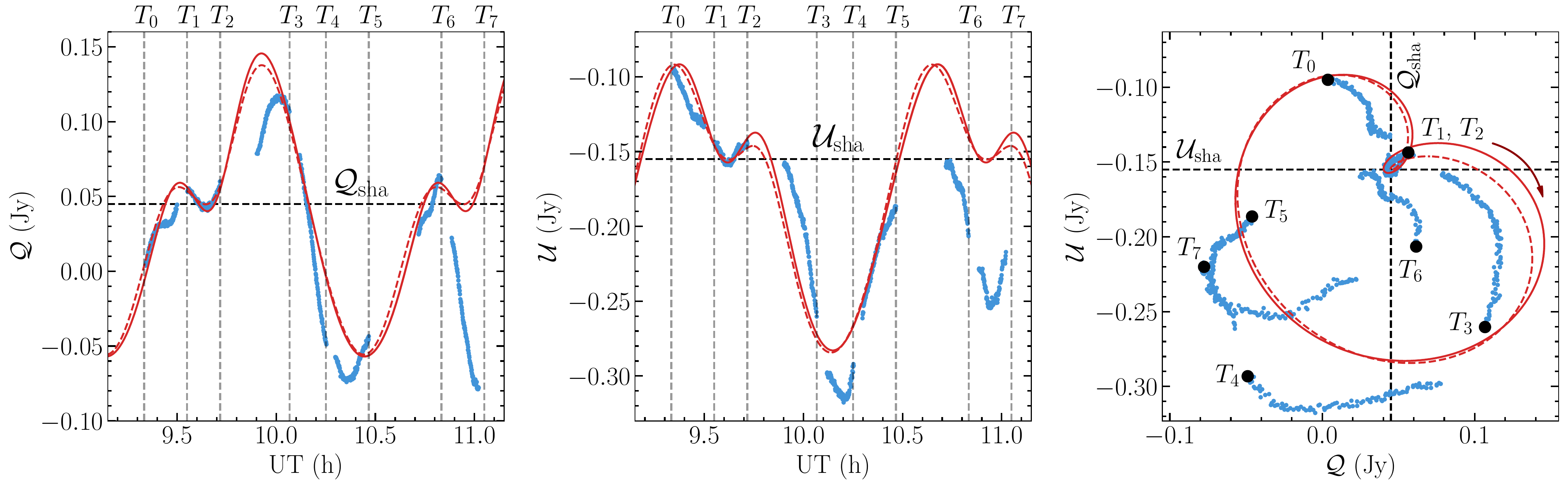}
    \caption{Comparison between ALMA data in the loopy period on 2017 Apr 11 (blue dots) and hot spot model prediction for the inclination $i=158^\circ$, the orbital radius $r_{\rm orb} = 11 M$, the vertical magnetic field, and the spin $a_*=0$. We compare predictions of a simplified semi-analytic model (dashed red lines, Appendix \ref{app:semi-analytic} and \citealt{Gelles2021}) with a slow light, full radiative transfer model \citep[continuous red line,][]{Vos2022}. A static, linearly polarized component of $|\mathcal{P}_{\rm sha}|=|\mathcal{Q}_{\rm sha}+i\, \mathcal{U}_{\rm sha}|=0.16$\,Jy and $\chi_{\rm sha} = -37^\circ$ was coherently added to the hot spot models, and the loop position angle was adjusted to match the observations.
    }
    \label{fig:data_model}
\end{figure*}

\subsection{Constraints on the system geometry}

Unlike GRAVITY, phased ALMA lacks the resolving power to constrain the brightness centroid motion coinciding with the polarimetric loops. Hence, to break the degeneracy between the orbital period and radius, we assume that the motion of the putative hot spot is Keplerian. Some discussions of the impact of Keplerianity on models and the interpretation of the observations are given in Appendix \ref{app:nonkepler}. While we generally expect the innermost flow to be sub-Keplerian \citep[e.g.,][]{Porth2021}, super-Keplerian pattern motion was also proposed as an interpretation of the GRAVITY observations \citep{Matsumoto2020}. The period of Keplerian orbits in Kerr spacetime is
\begin{equation}
  T(x) =  2 \pi t_{\rm g} \left(x^{3/2} + a_*\right) \ ,
\end{equation}
where $t_{\rm g} = GM_{\sbullet[1.35]}/c^3 = r_{\rm g}/c$, $x = r_{\rm orb}/r_{\rm g}$, and $a_*$ is the dimensionless black hole spin. Assuming $M_{\sbullet[1.35]} = 4.155 \times 10^6 M_\odot$, which is between the two competing results of \citet{Do2019} and \citet{Gravity2022}, we find $t_{\rm g}= 20.46$\,s. We estimate the period of the large loop shown in Fig.~\ref{fig:loops_data} to be $74\pm6$\,min. This range implies a Keplerian orbit radius $r_{\rm orb}$ in the $10.0-11.2\,r_{\rm g}$ 
 range for a Schwarzschild black hole ($a_* = 0$) or $9.8-11.0\,r_{\rm g}$ for a maximally spinning Kerr black hole ($a_* = 1$).

 We used a simplified semi-analytic model of an equatorial Keplerian hot spot developed by \citet{Gelles2021} to survey the parameter space of the orbital radius $r_{\rm orb}$, observer's inclination $i$, and magnetic field geometry $\boldsymbol{B} = [B_r,B_\phi, B_z]$, expressed in the cylindrical coordinate system tied to the hot spot orbital plane. Using the observables described in Section \ref{sec:loops_intro} we were able to constrain these parameters -- see the details provided in Appendix\,\ref{app:semi-analytic}. We concluded that only models with low inclination angle $i \sim\!20^\circ$ (or, equivalently, $i \sim\!160^\circ$) and dominance of the vertical magnetic field $B_z$ can reproduce the observed data features. These findings are remarkably consistent with the conclusions from the IR observations of hot spots by GRAVITY \citep{gravity_loops_2018,Baubock2020,Jimenez2020}.

\subsection{Direct comparison to data}

Having constrained the parameters of the geometrical hot spot model, we attempted a direct comparison with data using a more physical numerical modeling framework, described in detail in \citet{Vos2022}. Here we consider a Gaussian hot spot on an equatorial orbit around a Kerr black hole, simulated with the relativistic radiative transfer code \texttt{ipole} \citep{Moscibrodzka18}. We accounted for the finite velocity of light, hence the impact of the secondary images was correctly taken into account (see also Appendix \ref{app:appearance}). We performed a complete radiative transfer calculation, accounting for emission, self-absorption, internal Faraday rotation, and Faraday conversion, in order to compute the full $\mathcal{I,Q,U,V}$ Stokes vector of the observed emission. While we only discuss the LP here, comments on the total intensity and CP are given in Appendix\,\ref{app:StokesIV}.

The characteristic (but not unique) parameters that allowed us to match the model shown in Fig.\,\ref{fig:data_model} to the observed polarized flux density are the following: the number density of $n_{\rm e}\!=\!5\times 10^5\, {\rm cm}^{-3}$, the magnetic field $B\!=\!10$\,G, the dimensionless electron temperature of $\Theta_{\rm e} = k_{\rm B}T_{\rm e}/m_{\rm e}c^2\!=\!50$, and the Gaussian hot spot diameter of $\sim$ 6\,$r_{\rm g}$. 
For simplicity, we assumed a relativistic thermal distribution of the energy of electrons. We find that for such parameters, the system is both optically and Faraday thin. Given the turbulent, time-dependent character of the emission, we attempted to primarily reproduce the observational appearance of the small inner loop $A_{\rm tiny}$, as it is a complex feature that corresponds to a short time duration of 10\,min and hence is likely strongly dominated by the dynamical hot spot signature rather than by the stochastic, time-correlated variation of the observable shadow feature. In Fig.~\ref{fig:data_model} we impose the fiducial model with a spin $a_*\!=\!0.0$, Keplerian orbit at $r_{\rm orb}\!=\!11 r_{\rm g}$, the purely vertical magnetic field $B_z$, and the observer's inclination of 158$^\circ$ on the ALMA data. We adjusted the position angle (PA) of the $\mathcal{Q}$-$\mathcal{U}$ loop and a constant component $\mathcal{P}_{\rm sha}$ to approximately match the data features. We fixed the black hole spin to $a_* = 0$, but some hints of a larger spin are discussed in Appendix \ref{app:nonkepler}. The observed PA is corrupted by the Faraday rotation, and we discuss the intrinsic source orientation in Appendix\,\ref{app:orient}, where we estimate the PA of the hot spot spin axis projected on the sky to be $\sim 60^\circ$ east of north. We find a rather remarkable correspondence between the model and observations during the primary polarimetric loop ($T_0$-$T_5$). While more fine-tuned models could likely provide a better fit to data, the residuals can easily be explained with the corrupting effects and model limitations discussed in Section \ref{sec:limitations}.

\subsection{Limitations of the modeling}
\label{sec:limitations}

There are several important limitations to our modeling framework, which are listed in the order of relevance for the presented results:
 
 \begin{enumerate}
 
 \item Emission from the turbulent observable shadow component $\mathcal{P}_{\rm sha}$ is stochastically variable and correlated in time. With a standard deviation of $\sim\!0.08$\,Jy (see Fig.~\ref{fig:general_LC}), this component may contribute to the majority of the present discrepancies.
     
    \item We assume that hot spot emission is constant in the comoving frame.
The synchrotron cooling timescale for electrons is\footnote{The synchrotron cooling timescale for electrons $t_\mathrm{cool} \equiv u_{\rm e}/\Lambda$ where 
$u_{\rm e} = 3 n_{\rm e} \Theta_e m_{\rm e} c^2$ 
is the electron internal energy and $\Lambda \simeq 5.4 B^2 e^4 n_{\rm e} \Theta_{\rm e}^2 /(c^3 m_{\rm e}^2)$ is the synchrotron cooling rate for a thermal population of electrons with $\Theta_{\rm e} \gtrsim 1$ (see Appendix~A in \citealt{Moscibrodzka2011}).} $t_{\rm cool} \approx 11\,{\rm h} \times (B/10\,{\rm G})^{-2} \times (\Theta_{\rm e}/50)^{-1}$.
Hence, if the magnetic field is strong and the temperature is high, the hot spot may radiatively cool down appreciably on a timescale of $\sim\!1$\,h.
     This would increase the energetic output in the millimeter wavelengths toward the end of the observations (see also Appendix \ref{app:delays}). 
    
         \item Hot spots may be subject to shearing in a differentially rotating accretion flow \citep[e.g.,][]{Zamaninasab2010,Tiede2020}, which in turn causes depolarization. We expect that this affects the second polarimetric loop, between $T_5$ and $T_7$ in Fig.~\ref{fig:loops_data}, which is smaller and more distorted than the first one. For the shearing to be important on a relevant timescale, the hot spot must be relatively large, with a radius $r_{\rm hsp} > r_{\rm g}$.
    
     
          \item Fluctuations of the Faraday screen result in the time dependence of RM, as discussed in Appendix\,\ref{app:RM}. During the loopy period, a RM variation may contribute to the $(\mathcal{Q}_{\rm hsp},\mathcal{U}_{\rm hsp})$ vector PA by as much as 20$^\circ$, but the exact angle depends on the unknown details of the Faraday screen geometry.
     
          \item We assume a circular motion, but there could be a non-negligible radial component of the hot spot velocity, either following the infall of matter onto a black hole, or perhaps even as an ejection of a magnetic flux tube from the system \citep{Dexter2020,Porth2021}.

          \item We assume a spherically symmetric Gaussian structure of the hot spot, while in reality it is more likely an irregular region of the accretion flow (see, e.g., \citealt{Ripperda2022}). This impacts how the polarized emission integrated over the hot spot volume decorrelates with the spatially variable magnetic field.

     \item We only consider a hot spot moving in the equatorial plane, but the orbit could be displaced vertically, and the observed feature could, in principle, be located in the jet stream \citep{Ball2021}.
     
         \item The magnetic field structure may be nonaxisymmetric or time-dependent, and itself subject to the turbulence of the accretion flow.

 \end{enumerate}
Because of these important limitations of the model specification and applicability translating into poorly characterized systematics, we did not attempt a formal fit to the data. However, with a reasonably general model, accounting for the time-correlated corruption effects, this could be a fruitful path forward.

\section{Summary and discussion}
\label{sec:discussion}

A hot spot orbiting black hole in a clockwise direction in the sky, observed at low inclination, is a simple scenario which explains the linear polarization of ALMA millimeter light curves of flaring \sgra remarkably well. Models with vertical magnetic fields threading the hot spot represent data characteristics far better than those with purely toroidal or radial magnetic fields. While in this paper the model is postulated for a single observed event, the hot spot interpretation of millimeter light curves' variability following X-ray flares will be tested through future observations of \sgra.

In this paper, we adopted simple geometric models to interpret observations.
A more physically complete model can be developed using the time-dependent general relativistic magnetohydrodynamic simulations (GRMHD) of radiatively inefficient accretion flows representing magnetically arrested disks \citep[MADs;][]{Narayan2003,Dexter2020}. In such systems magnetic reconnection episodes occurring near the black hole event horizon can expel magnetic flux in large eruptions. For example, \citet{Ripperda2022} reported orbiting hot spots in the inner 10-30\,$r_{\rm g}$, corresponding to flux tubes of a vertical magnetic field, produced by the reconnection.
The flux tubes remain coherent for $\sim$1 orbit and confine hot, low-density plasma, in comparison to the surrounding colder, higher-density accretion disk with a mostly toroidal magnetic field. Individual plasmoids are less suitable as a model for the observed $\mathcal{Q}$-$\mathcal{U}$ loops since they remain smaller than 1\,$r_{\rm g}$, have a higher density, and originate closer to the black hole event horizon. Additionally, individual plasmoids in no-guide-field reconnection, as occurs near the horizon, mainly contain a helical magnetic field \citep[e.g,][]{Ripperda2020,Nathanail21}. 
We note, however, that the temporal behavior and hot spot formation in the GRMHD models is stochastic and may depend on resolution or initial plasma conditions and magnetic field configuration, so finding the solution matching \sgra data quantitatively is challenging. 
Hence, in this paper we do not attempt to model the source with GRMHD simulations, but we hope that the presented simplistic model will guide more complex work in the future.
 
The EHT Collaboration published VLBI images of \sgra total intensity, taken on 2017 Apr 6 and 7 \citep{SgraP1,SgraP3}. Comparisons with the GRMHD models revealed a preference for viewing angles of $i\le30^\circ$ \citep{SgraP5}, in agreement with inclination inferred from our polarimetric ALMA data. The direction of rotation in the EHT images is currently unconstrained, but the analysis presented in this paper suggests that the ring in \sgra should be rotating in the clockwise direction in the sky, assuming that its direction should be consistent with the inferred rotation of the hot spot.
Additionally, GRMHD MAD models with significant vertical magnetic fields are preferred in the EHT analysis. 
Nonimaging, proto-EHT polarimetry of \sgra also points toward organized magnetic fields on the relevant scales \citep{Johnson2015science}.
We conclude that it is very probable that the X-ray flare and the subsequent millimeter bright spot produced in \sgra on 2017 Apr~11 were generated by magnetic reconnection within the magnetically arrested accretion flow. 
In the future, polarimetric \eht images and movies of \sgra should give us clearer information on the geometry and evolution of magnetic fields near the event horizon of the black hole; for example, readers can see how polarimetric images of event horizon scale emission in M87$^*$ in \citealt{Paper7} are strongly constraining for GRMHD simulations in \citealt{Paper8}.

The models that we employed are similar to the ones used to interpret IR observations of \sgra by the GRAVITY instrument, and so are the inferred parameters of the observed system \citep{gravity_loops_2018, Baubock2020,Jimenez2020}. The two instruments may be observing the same phenomenon, perhaps with millimeter polarimetric loops appearing with a delay, after a hot spot cools down. There are, however, interesting discrepancies. IR observations suggest shorter periods, and a brightness centroid loop shifted from the gravity center, which may be an indication of a nonequatorial motion in the base of the jet \citep[e.g.,][]{Ball2021}. Future simultaneous multiwavelength observations will play an essential role in revealing the relationship between orbiting hot spots observed at different frequencies.

\section*{Acknowledgements}
{We thank M. Bauböck, A. Broderick, K. Chatterjee, J. Dexter, F. Eisenhauer, S. von Fellenberg, D. Haggard, E. Himwich, S. Issaoun, A. Jim\'enez-Rosales, M. Johnson, D. Marrone, D. Palumbo, D. Pesce, B. Ripperda, P. Tiede, M. Zaja\v{c}ek, and the EHT Polarimetry Working Group for helpful comments, discussions, and encouragements. This paper makes use of the following ALMA data: ADS/JAO.ALMA\#2017.1.00797.V. ALMA is a partnership of ESO (representing its member states), NSF (USA) and NINS (Japan), together with NRC (Canada), NSC and ASIAA (Taiwan), and KASI (Republic of Korea), in cooperation with the Republic of Chile. The Joint ALMA Observatory is operated by ESO, AUI/NRAO and NAOJ. This work was supported by the Black Hole Initiative at Harvard University, which is funded by grants the John Templeton Foundation and the Gordon and Betty Moore Foundation to Harvard University. MM acknowledges support by the Dutch Research Council (NWO) grant No. OCENW.KLEIN.113. This work has been partially supported by the Generalitat Valenciana GenT Project CIDEGENT/2018/021 and by the MICINN Research Project PID2019-108995GB-C22. We also thank A. Elbakyan for her contributions to the open science initiative.}

\bibliography{sgra_polar}
\bibliographystyle{aa}

\clearpage

\begin{appendix}

\addtolength{\parskip}{2mm}


\section{Variable rotation measure}
\label{app:RM}

\begin{figure}[h!]
    \centering
    \includegraphics[width=\columnwidth]{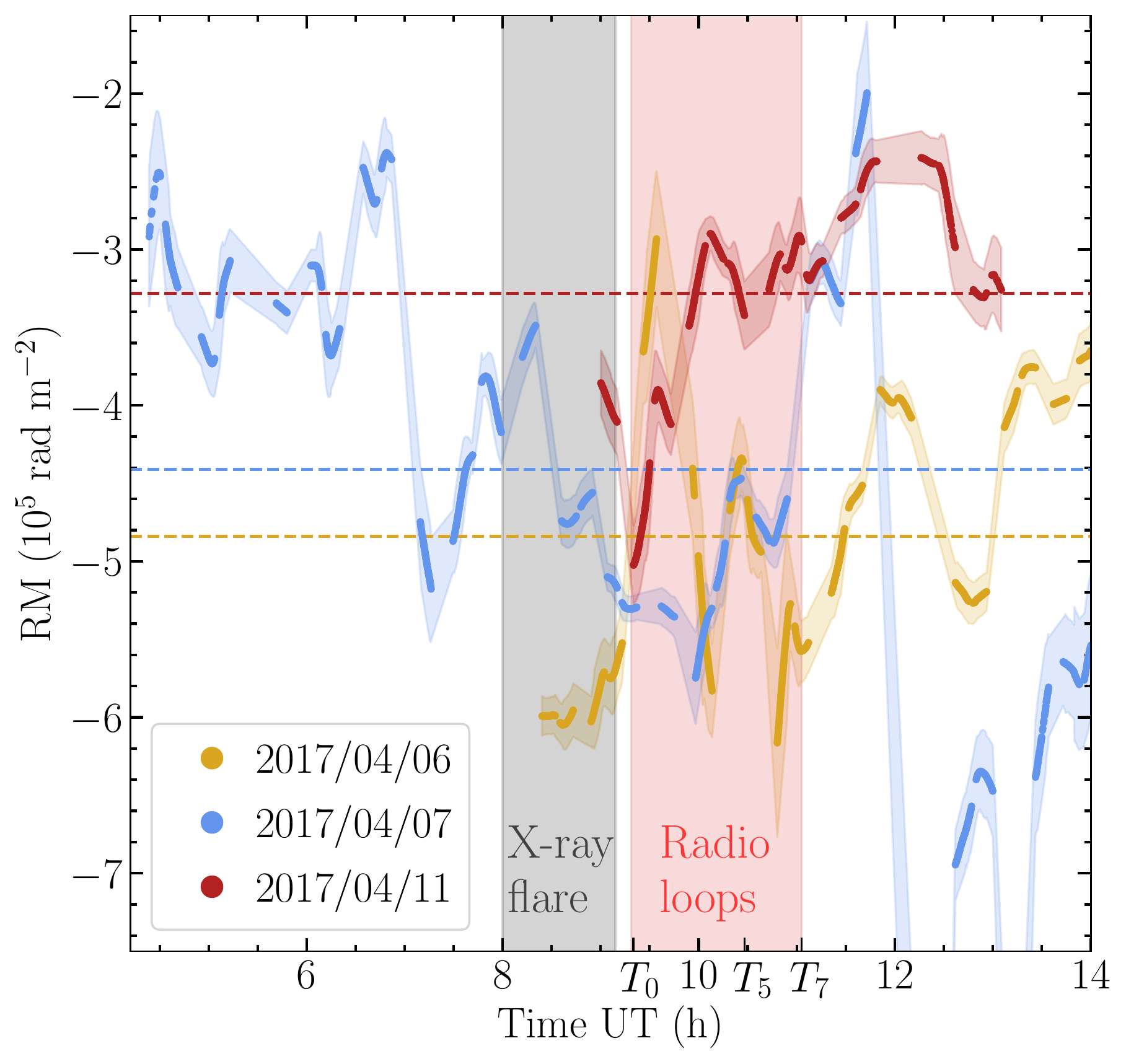}
    \caption{Rapid variation of the RM at millimeter wavelengths in \sgra. The plot corresponds to the RM fit with linear regression across all four bands on 4\,s cadence, smoothed with a Gaussian filter with 6\,min standard deviation, with color bands around the mean RM indicating the local signal standard deviation. Dashed lines represent the mean daily values reported by \citet{Goddi2021}. We observe rapid swings of the RM on timescales $\sim$0.5\,h, indicating a presence of the intrinsic Faraday screen component corresponding to the innermost region of the accretion flow. The vertical gray and red bands indicate the time range of the X-ray flare and polarimetric loops observed on 2017 Apr 11.}
    \label{fig:RM}
\end{figure}

Polarized radiation emitted by the compact \sgra source is affected by the Faraday effect. Under simplifying assumptions of an external character of the Faraday screen, it causes rotation of the EVPA by the angle proportional to the square of the wavelength with respect to the intrinsic (zero wavelength) EVPA $\chi_0$,
\begin{equation}
    \chi = \chi_0 + \text{RM} \, \lambda^2 \ .
\end{equation}
The proportionality constant RM is the rotation measure, which is dependent on the gas density, magnetic field along the line of sight, and the plasma temperature \citep[if the plasma is relativistically hot;][]{Quataert2000}. We confirm the previously reported value of RM at
$\sim\!220$ GHz, to be $\approx - 5 \times 10^5\,{\rm rad/m^2}$ (e.g., \citealt{Bower2018}), with our measurement of $(-4.3 \pm 1.3)\times 10^5\,{\rm rad/m^2}$ corresponding to the mean of the entire observing campaign.
Given the uniquely high cadence and S/N of our observations, we can study the RM variability on short timescales,
elaborating on the results of \citet{Goddi2021}, where single values per day were provided for the same data set. As
shown in Fig.~\ref{fig:RM}, it is not unusual for the RM to change
by $\sim\!1 \times 10^5\,{\rm rad/m^2}$ within $\sim$\,30 min, which is a dynamical timescale at the innermost stable circular orbit (ISCO) of \sgra modeled as a nonspinning black hole. Moreover, the intra-day variability of RM is generally larger than the variability between the three observing days. This implies a large
intrinsic component of the Faraday screen, most likely corresponding to
the rapidly fluctuating gas density 
in a turbulent accretion flow very close to the black hole event horizon. We note that future observations of hot spots, performed by ALMA at higher available frequency, could greatly reduce systematic uncertainties related to the RM variation. 

Persistence of the RM sign, along with the persistence of the sign of the circular polarization and the fractional LP magnitude (see Section~\ref{sec:mean_lc_properties}), suggest that there is a structured magnetic field of  well-defined geometry present in the \sgra system, which does not change dramatically with the accretion flow turbulence. The presence of such dynamically important magnetic fields near the black hole event horizon is a characteristic theoretical expectation from MAD systems \citep{Narayan2003,SgraP5}.

\section{Semi-analytic model constraints}
\label{app:semi-analytic}

\begin{figure*}[t]
    \centering
    \includegraphics[width=\linewidth]{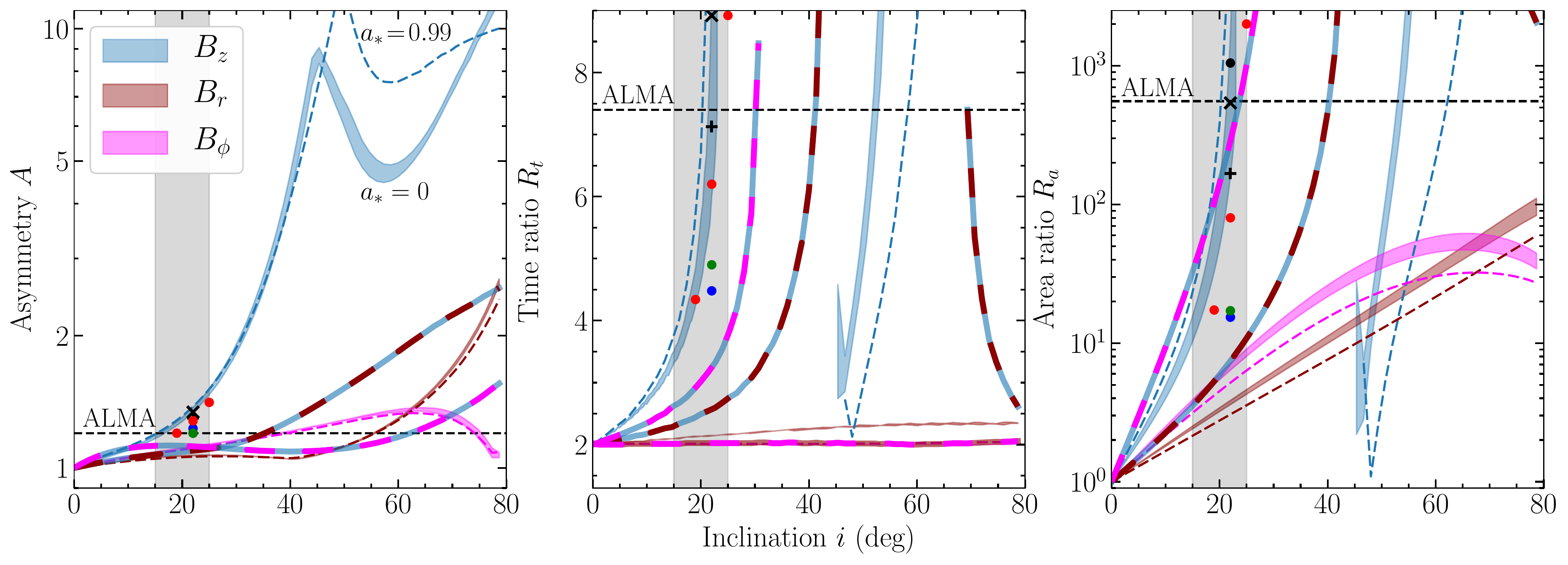}
    \caption{Three observables constrained with ALMA (dashed horizontal lines), evaluated for the semi-analytic model of \citet{Gelles2021}. The thick, semi-transparent lines correspond to Keplerian models with a zero spin and orbital radius ranging from $10\,r_{\rm g}$ to $11\,r_{\rm g}$ and a fixed magnetic field orientation, following the legend. Additionally, we show the Keplerian models with a spin of $a_*=0.99$ and $r_{\rm orb} = 11\,r_{\rm g}$ with dashed lines, and two cases of mixed magnetic fields with a spin of $a_*=0.0$ and $r_{\rm orb} = 11\,r_{\rm g}$: $B_z = B_r, B_\phi=0$ (dashed blue and dark red) and $B_z = B_\phi, B_r = 0$ (dashed blue and magenta). We also show predictions of the numerical model of  \citet{Vos2022}, the fiducial model shown in Fig.~\ref{fig:data_model} (red circles), and several models deviating from the fiducial one by changing a single parameter: a sub-Keplerian orbit at $r_{\rm orb} = 8\,r_{\rm g}$ (black circles), super-Keplerian orbit at $r_{\rm orb} = 14\,r_{\rm g}$ (green circles), a combined vertical and radial magnetic field (blue circle), a spin of $a_* = 0.99$ (black "$\times$"), and a spin of $a_* = 0.5$ (black "+").
    }
    \label{fig:constraints}
\end{figure*}

The most elementary model of the hot spot emission corresponds to a synchrotron point source in an equatorial Keplerian (circular geodesic) orbit of radius $r_{\rm orb}$, moving through a static axisymmetric magnetic field $\boldsymbol{B} = [B_r,B_\phi, B_z]$ in vacuum, viewed by a distant observer at an inclination angle $i$. In such a case, null geodesics in Kerr spacetime can be readily computed semi-analytically \citep[e.g.,][]{Gralla2020}, with LP being evaluated following the conservation of the Penrose-Walker constant \citep{Walker1970}. Such a simplified model, implemented by \citet{Gelles2021}, ignores the effects related to the radiative transfer, the optical and Faraday depth, the finite size of the hot spot, as well as those related to the finite time of light propagation -- it employs a fast light approximation, implying that the secondary images, emerging from photons looping around the black hole in a strongly curved spacetime \citep{Darwin1959} are not treated correctly\footnote{A procedure to account for the time delay in the zero spin case with a single numerical time integral is discussed by \citet{Gelles2021}, but was not used here.}. Since these effects are expected to be subdominant, we nevertheless used the semi-analytic model to compare its predictions with the observations, and to obtain general constraints on the system geometry, before verifying our findings with a more physical model of \citet{Vos2022}. A direct comparison between the models of \citet{Gelles2021} and \citet{Vos2022} is given in Fig.~\ref{fig:data_model}, showing a high degree of consistency for the considered low Faraday depth example. A clear benefit of the semi-analytic model is that it is very computationally cheap, allowing us to survey a broad parameter space efficiently.

We employed the model of \citet{Gelles2021} to compare its predictions to observables described in Section~\ref{sec:loops_intro}. None of the considered quantities $(A,R_t,R_a)$ distinguish between the inclination of $i$ and $180^\circ - i$, or between the clockwise and counterclockwise rotation of the hot spot. Our findings are summarized in Fig.~\ref{fig:constraints}. In the first panel, we quantify how the small observed degree of asymmetry of the $\mathcal{Q}$-$\mathcal{U}$ loop generally disfavors large inclination, particularly for poloidal magnetic field configurations with $B_\phi = 0$. As noticed by \citet{gravity_loops_2018}, a purely toroidal (azimuthal) magnetic field implies a time ratio between the total period (the hot spot orbital period) and the inner loop of $R_t = 2$. This is a consequence of the origin of the inner loop in such a configuration being related to the Doppler deboost. Hence, our observation of $R_t \sim\!7$ rules out the dominance of the azimuthal magnetic field and implies that a poloidal magnetic field component is needed -- see the middle panel of Fig.~\ref{fig:constraints}. With a vertical magnetic field and low inclination, the inner loop originates from geometry between the magnetic field and the line of sight (subject to lensing), rather than from a Doppler effect, see also Appendix~\ref{app:appearance} and \citet{Vos2022}. Similar conclusions follow from the loop area ratio $R_a$ measurement. The impact of the black hole spin is already strongly subdominant at $r_{\rm orb} \sim\!10\,r_{\rm g}$, particularly at low inclinations, for which trajectories of photons do not approach the event horizon, see also Appendix~\ref{app:nonkepler}. If we put these constraints together, we conclude that only models with a low inclination angle $i \sim\!20^\circ$ (or, equivalently, $i \sim\!160^\circ$)  and dominance of the vertical magnetic field $B_z$ can reproduce the observed properties of the polarimetric loop. 
We notice, however, that a configuration with a slightly higher inclination and added nonvertical magnetic field component could also reproduce the observed properties -- a broader exploration of the parameter space would be most helpful here. 


\section{Impact of other model parameters}
\label{app:nonkepler}

\begin{figure*}[t]
    \centering
    \includegraphics[width=0.33\linewidth]{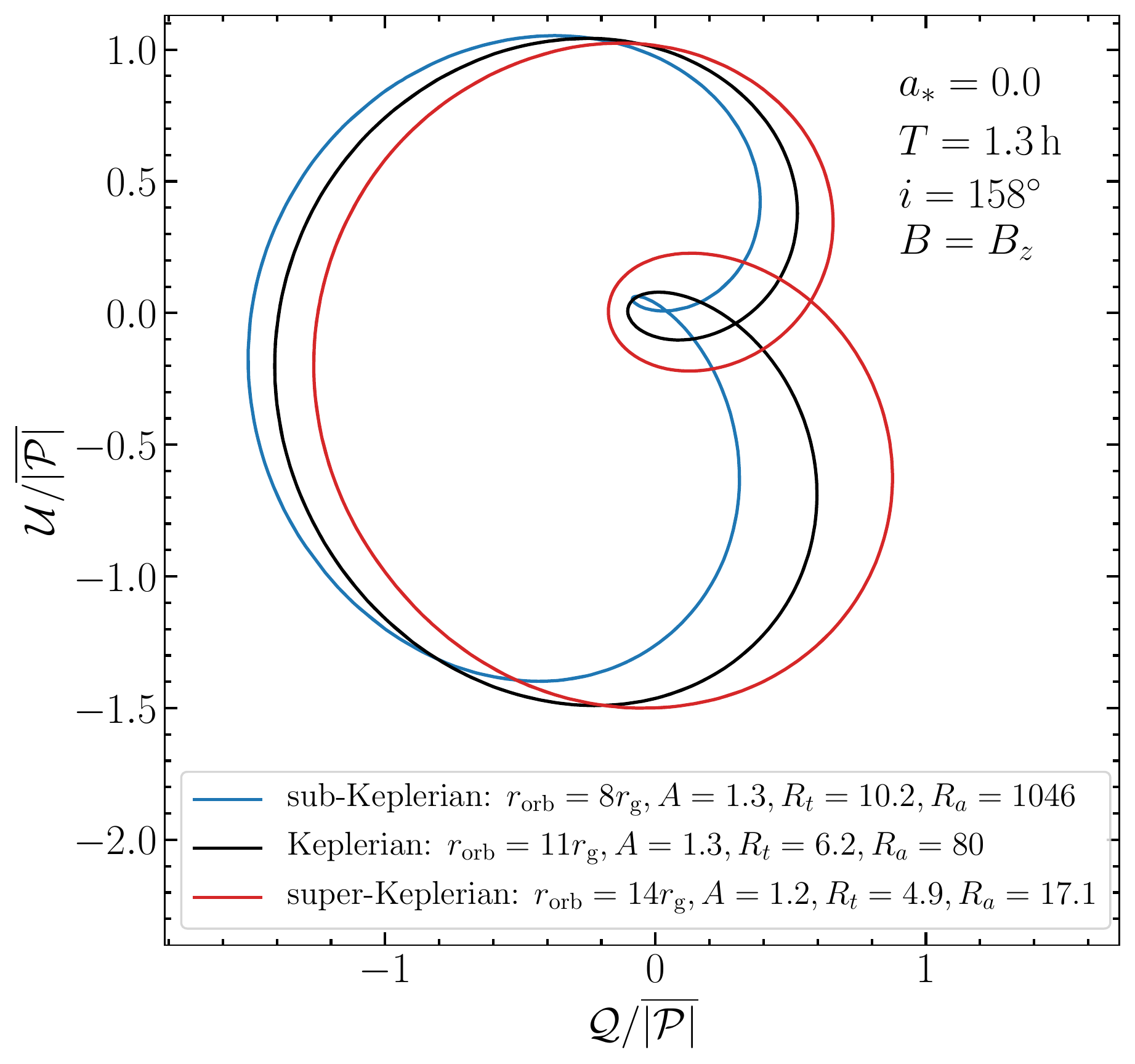}
    \includegraphics[width=0.33\linewidth]{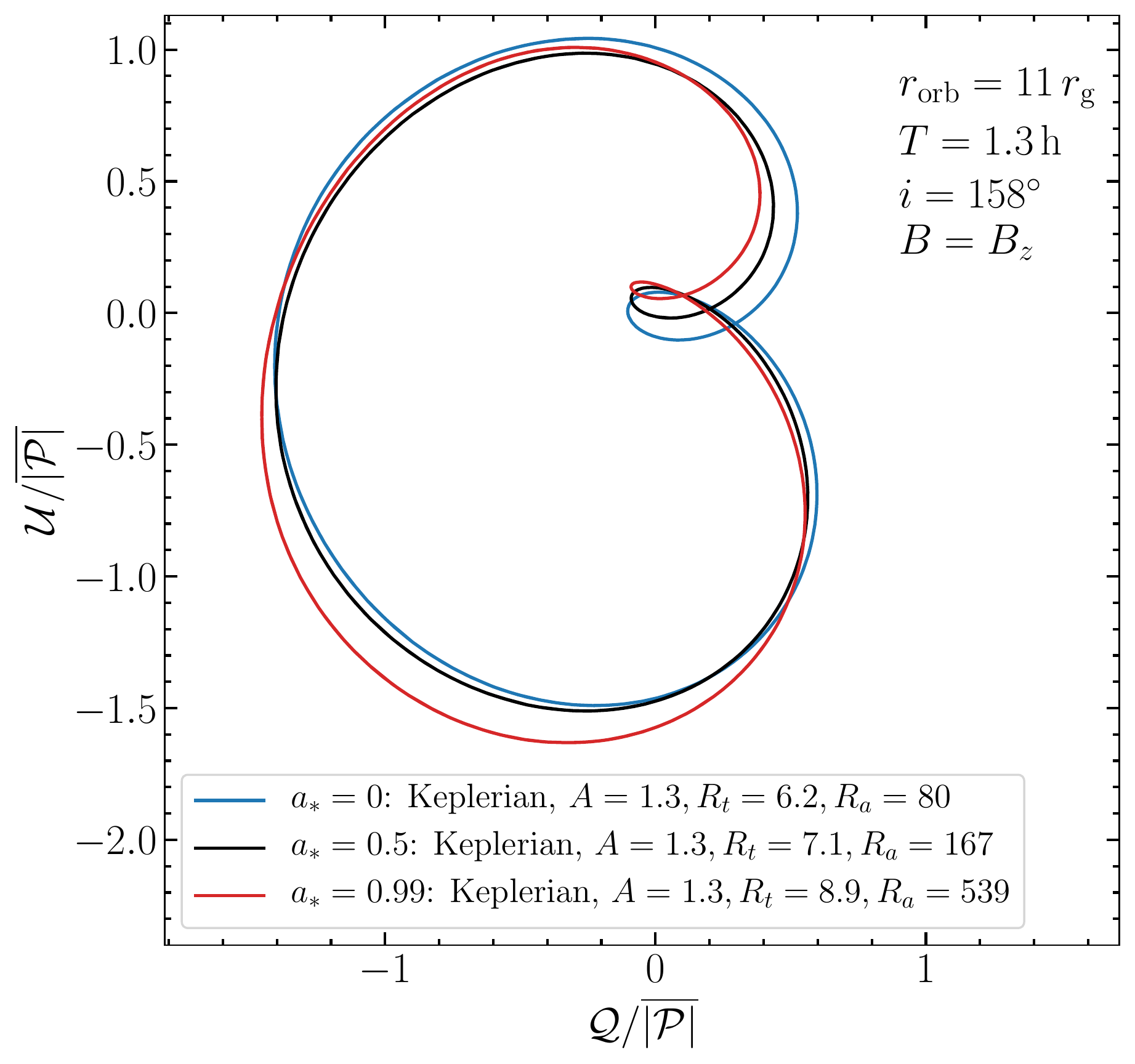}
    \includegraphics[width=0.33\linewidth]{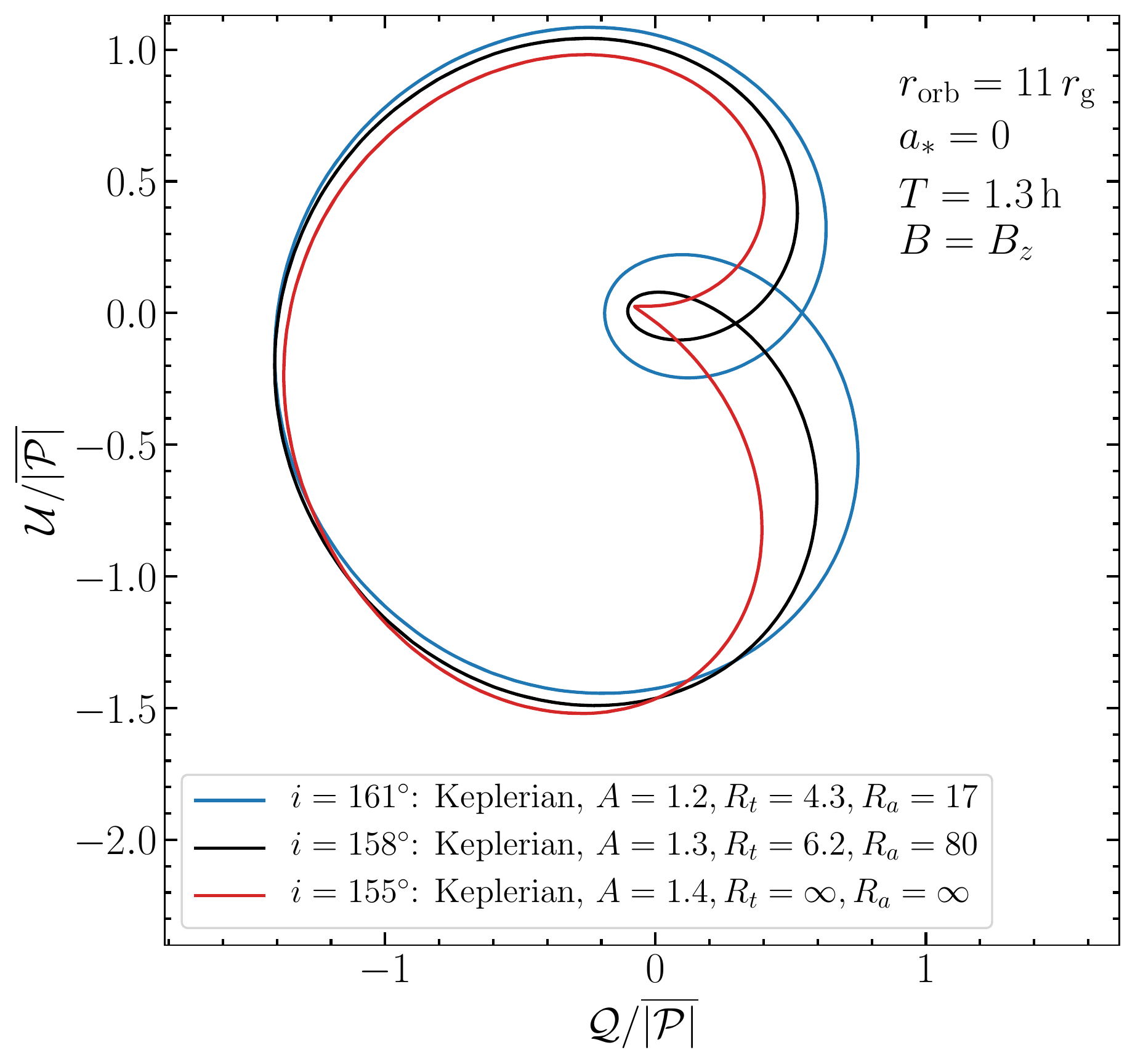}
    \caption{Impact of modifying parameters of the fiducial model: Keplerianity and orbital radius with a fixed orbital period (\textit{left}), black hole spin (\textit{center}), and inclination (\textit{right}). The presence and size of the small inner loop puts strong constraints on the inclination and orbital radius admitted by the vertical magnetic field model. Additionally, in the vertical magnetic field model framework, the observed loop size may suggest a large positive spin of \sgra.}
    \label{fig:nonkepler}
\end{figure*}

For the low inclination models dominated by the vertical magnetic field, which we identified as ones that are capable of matching the features observed by ALMA, we found that the inner loop becomes smaller ($R_a$ increases) with increasing spin, smaller with growing inclination (i.e., away from $0^\circ$ or $180^\circ$), smaller with a growing radius of a Keplerian orbit, larger ($R_a$ decreases) with a growing nonvertical component of the magnetic field, and larger for more super-Keplerian orbits of the same period. At the same time, the loop asymmetry $A$ increases with spin and inclination, but decreases with a nonvertical magnetic field and for more super-Keplerian motion of the same period. Some of these dependencies are illustrated in Fig.~\ref{fig:nonkepler} for the pretzel-like $\mathcal{Q}$-$\mathcal{U}$ loop corresponding to the fiducial hot spot model. 

The observations suggest a $\mathcal{Q}$-$\mathcal{U}$ loop asymmetry slightly lower, and the inner loop relatively smaller than the fiducial model parameters. In Fig.~\ref{fig:nonkepler} we observe that for a vertical magnetic field configuration ($B=B_z$) and low optical depth, loop parameters are very sensitive to Keplerianity and inclination. Increasing the orbital radius to $r_{\rm orb} = 14\,r_{\rm g}$ (super-Keplerian model) or changing inclination from  $i = 158^\circ$ to $i=161^\circ$ renders the inner loop far too large. Decreasing $r_{\rm orb}$ to $8\,r_{\rm g}$ (sub-Keplerian model) or changing inclination from  $i = 158^\circ$ to $i=155^\circ$ is a sufficient change to make the inner loop extremely small or to destroy it altogether. These considerations can be used to obtain more definite limits on Keplerianity and inclination in the framework of our model, thus we can claim $r_{\rm orb} = 11\pm 3 r_{\rm g}$ and $i = 158 \pm 3$\,deg with a degree of confidence. At the same time, a relatively smaller observed inner loop (larger $R_a$ and $R_t$) than for a fiducial Schwarzschild black hole model may be interpreted as a hint of a larger positive black hole spin, that is, a prograde motion of the hot spot with respect to the black hole rotation. This suggestion is further supported by our observation that deviating the magnetic field from purely vertical configuration generally also increases the inner loop size. Nevertheless, obtaining strong conclusions require more systematic inspection of the parameter space and, in any case, would be rather speculative at this time given the large systematic model uncertainties. For example, modifying the magnetic field to include a radial or azimuthal component while simultaneously increasing the model inclination may yield a similar effect as the spin increase, see also Fig.~\ref{fig:constraints}. Regardless of the current difficulties, it should be stressed that aggregating constraints from multiple observed hot spots, or obtaining additional constraints on the inclination and the magnetic field geometry, could lead to a robust measurement of the \sgra black hole spin in the future through the analysis of the $\mathcal{Q}$-$\mathcal{U}$ loops' shape.

\section{Appearance of the resolved model}
\label{app:appearance}

\begin{figure*}[t]
    \centering
\includegraphics[trim={1mm -7mm 0 2mm},clip, width=0.245\linewidth]{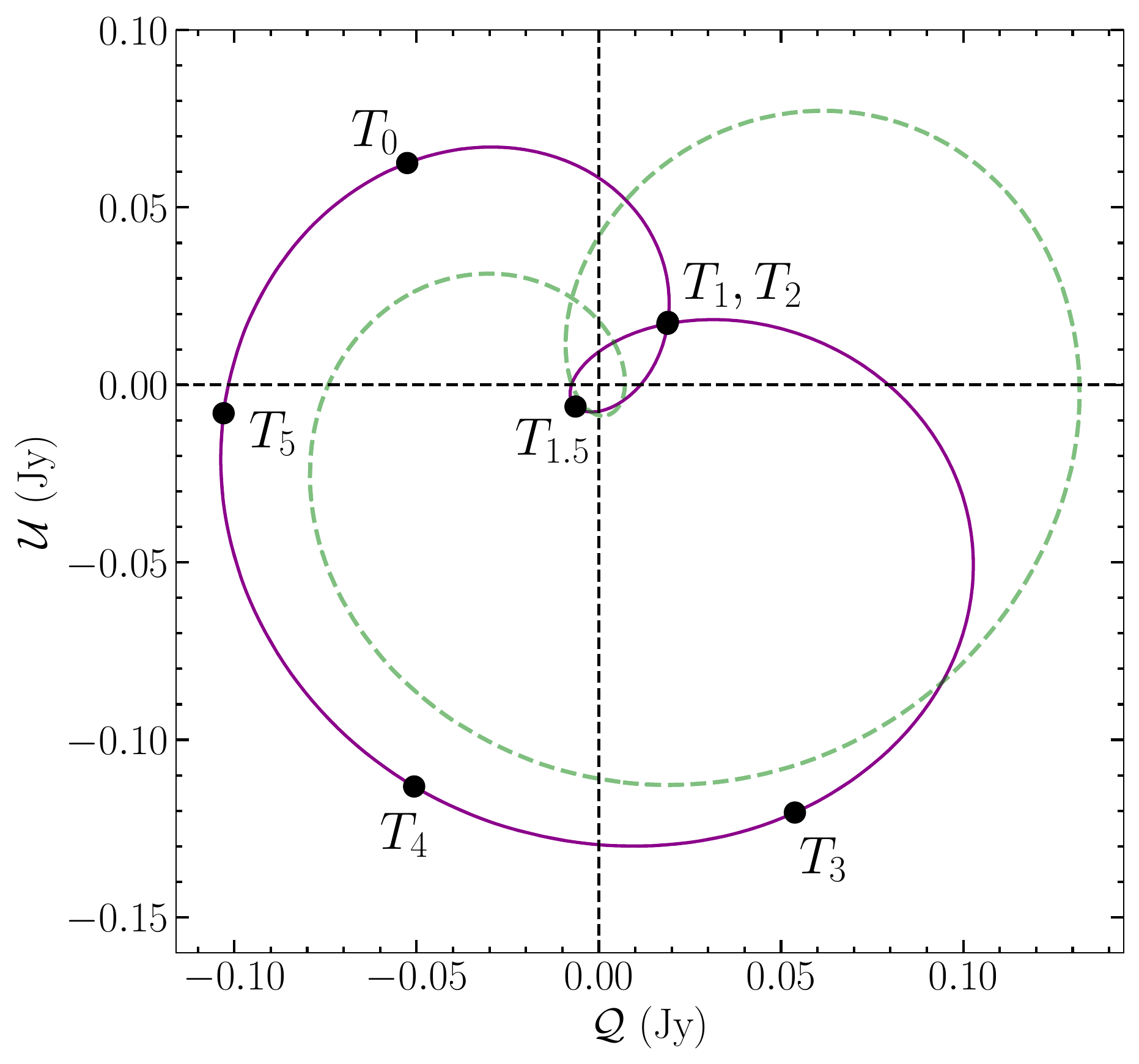}
\includegraphics[width=0.245\linewidth]{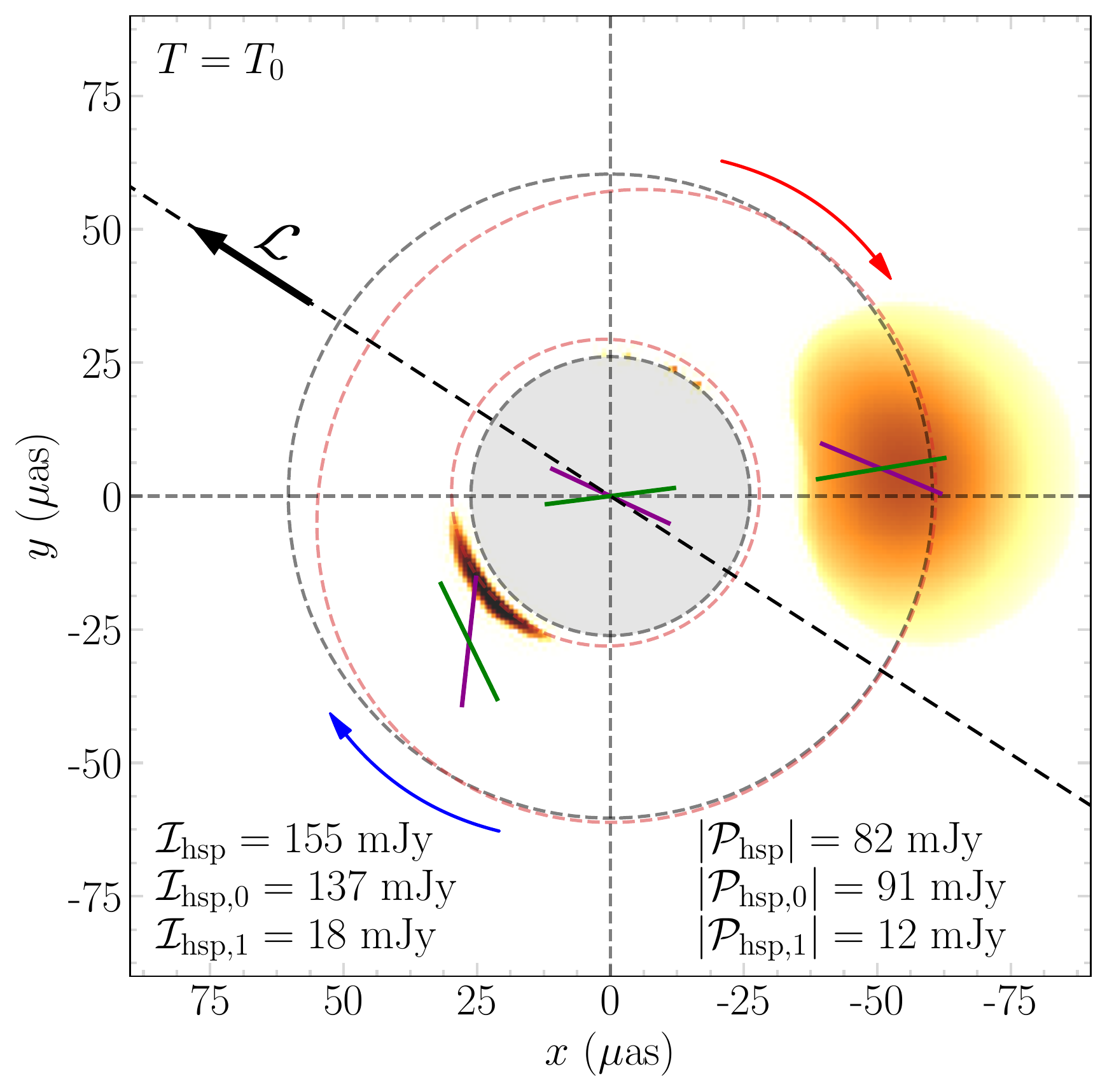}
\includegraphics[width=0.245\linewidth]{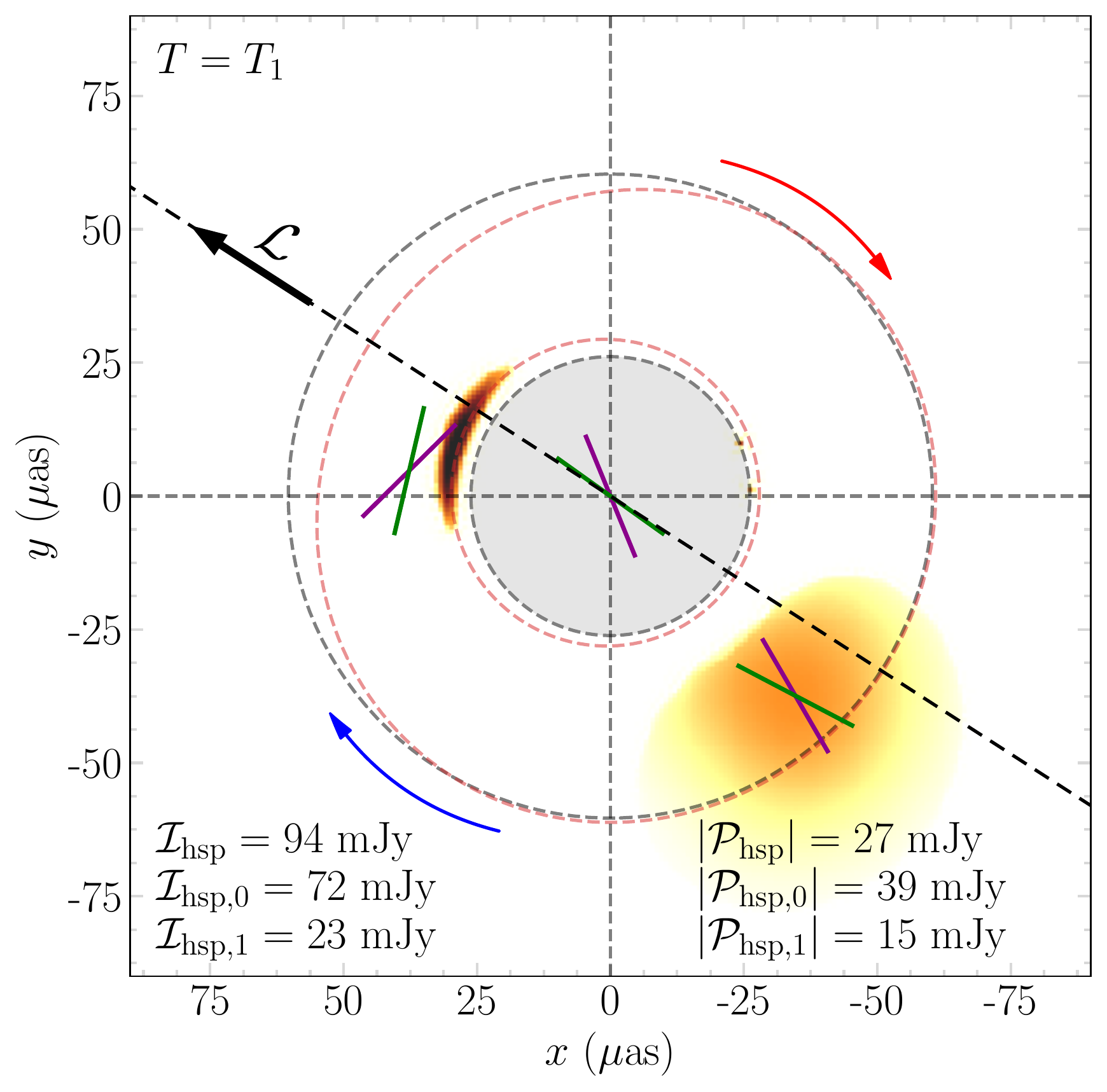}
\includegraphics[width=0.245\linewidth]{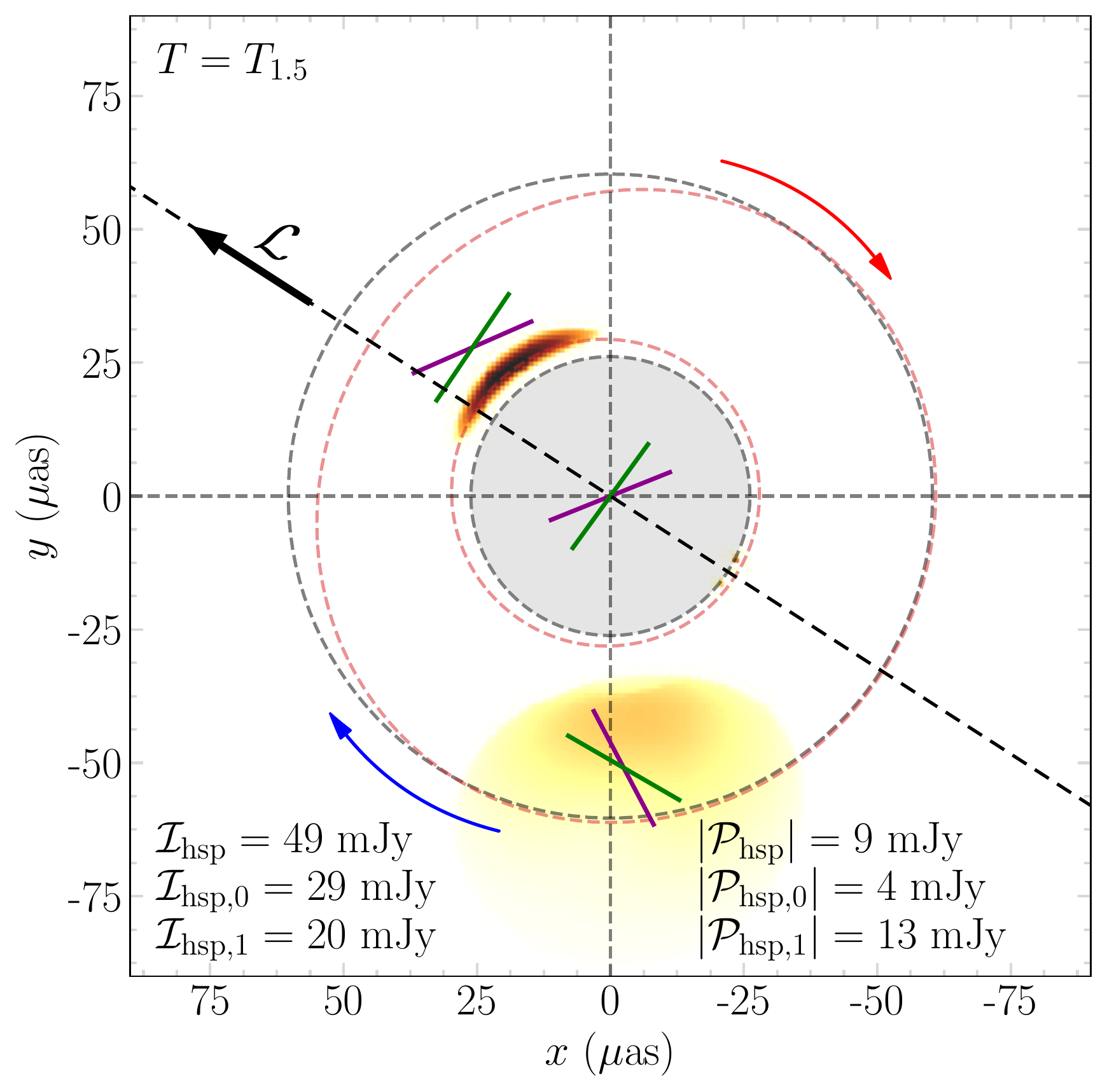}
\includegraphics[width=0.245\linewidth]{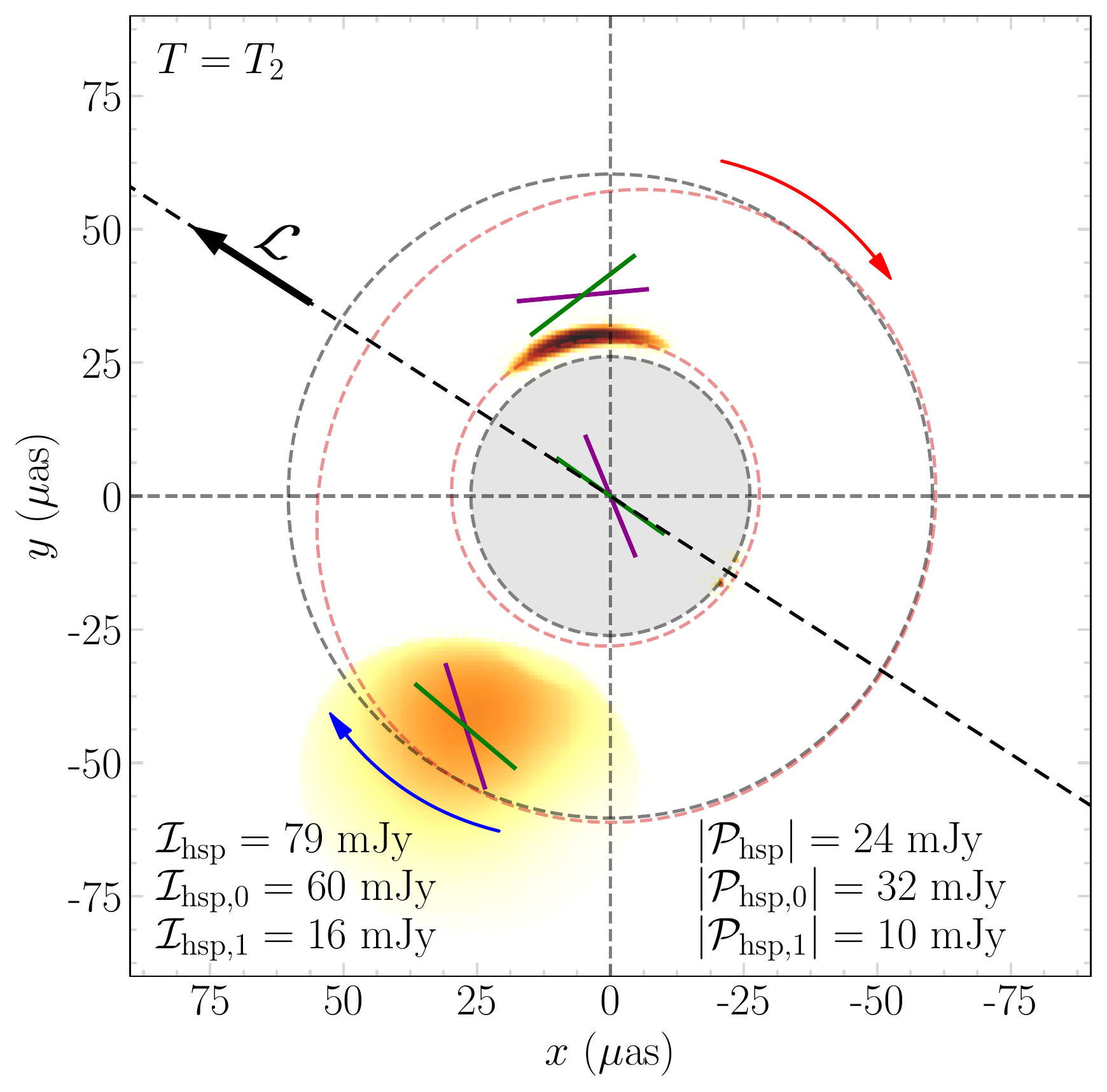}
\includegraphics[width=0.245\linewidth]{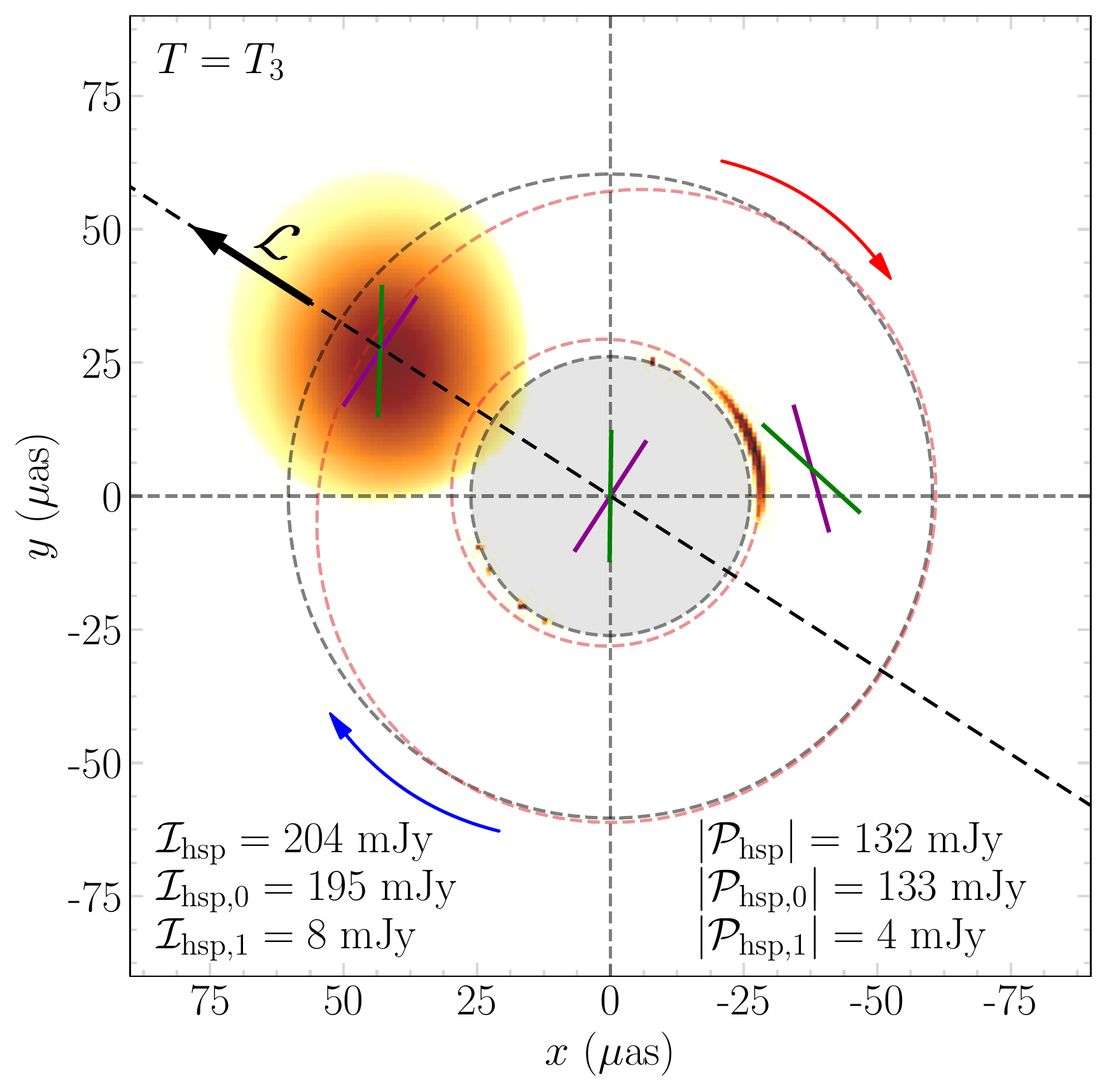}
\includegraphics[width=0.245\linewidth]{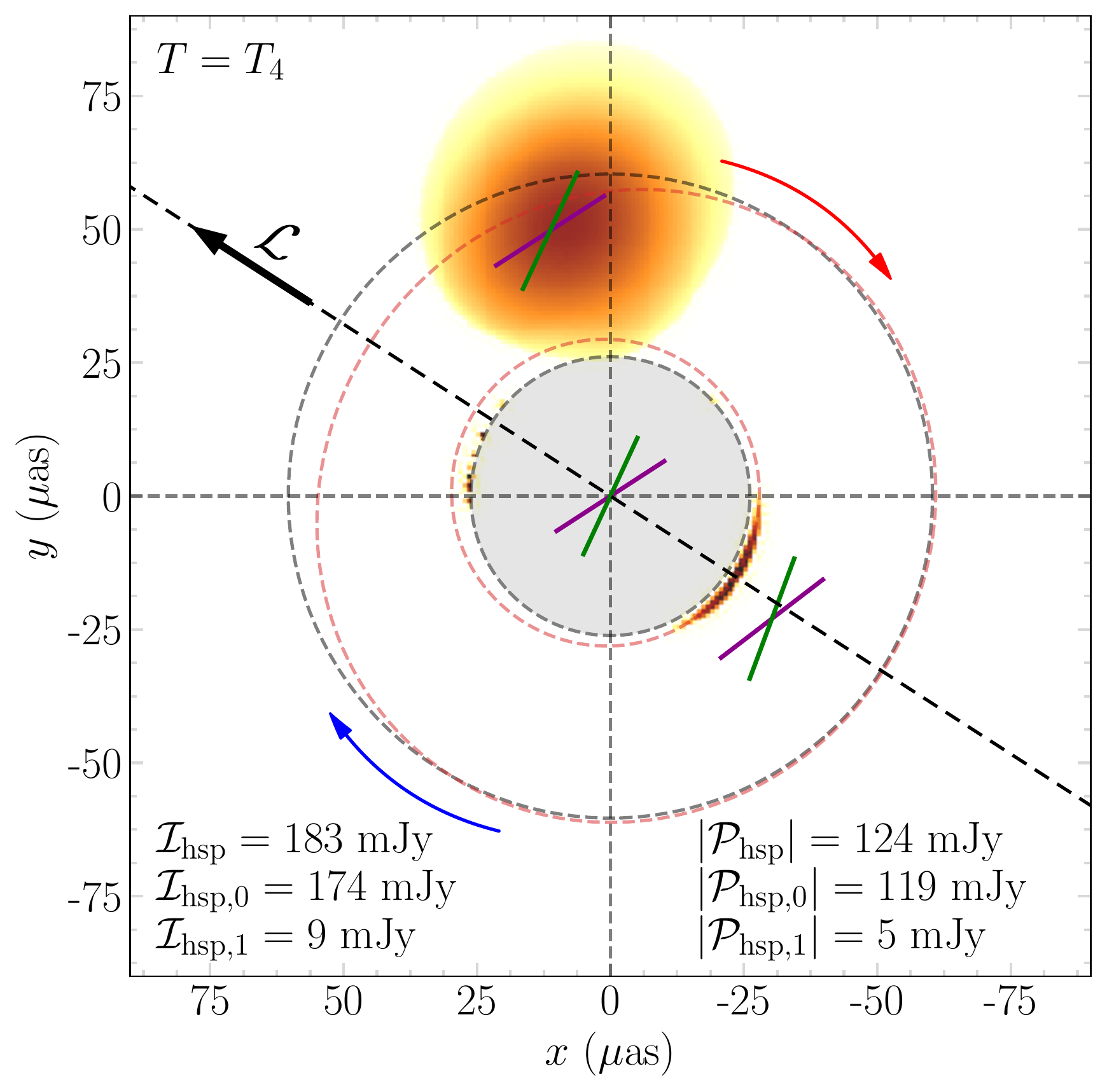}
\includegraphics[width=0.245\linewidth]{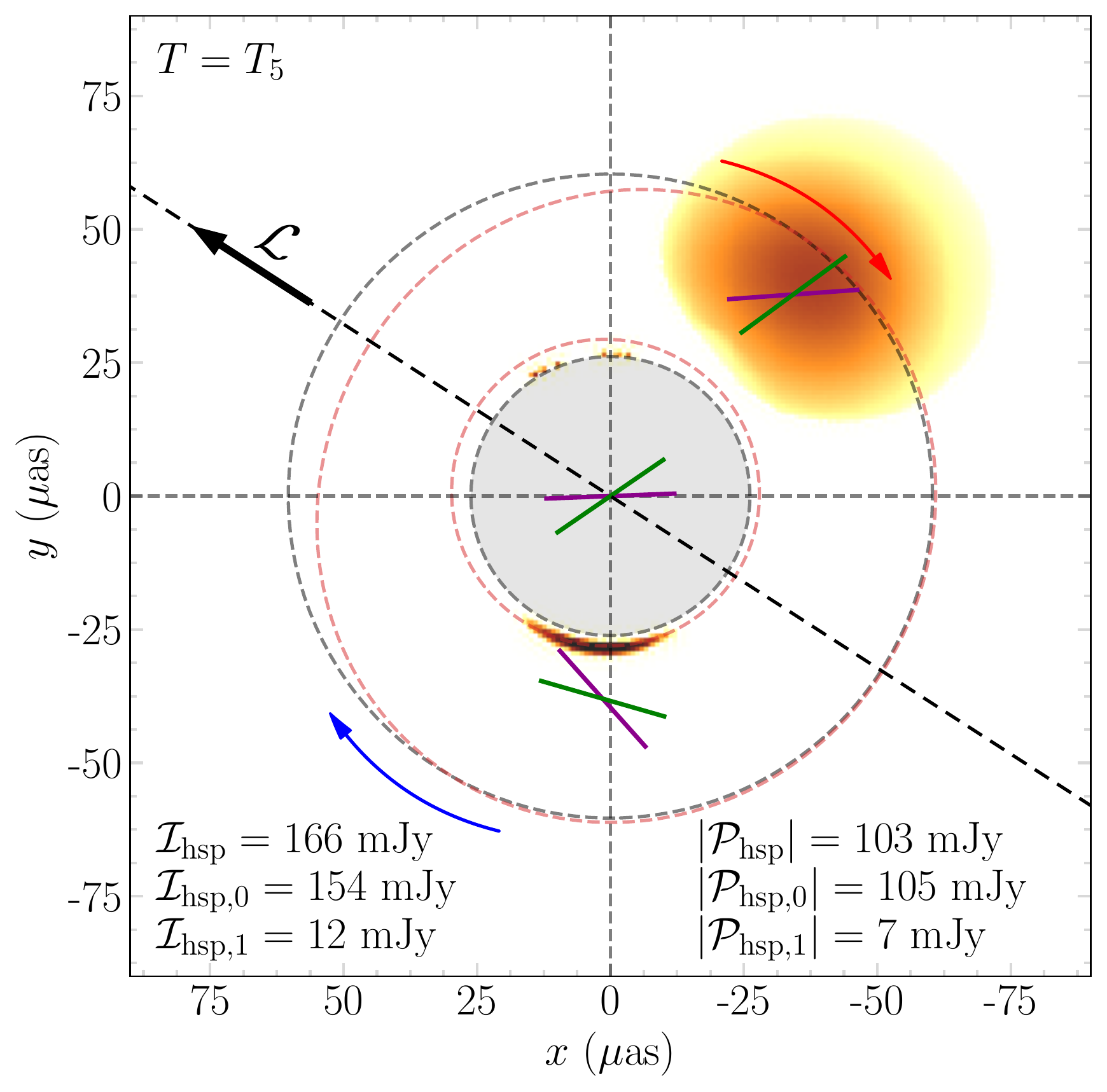}
    \caption{Spatially resolved appearance of the fiducial Keplerian model at times $T_1-T_5$ matching observed times defined in Fig.~\ref{fig:loops_data}. The first panel shows the prediction for the observed $\mathcal{Q}$-$\mathcal{U}$ loop shape with locations of individual snapshots being indicated (dark magenta line), as well as the intrinsic loop shape, after correcting for the Faraday rotation (dashed green line; see Appendix\,\ref{app:orient}). The remaining panels show the resolved system geometry at times $T_1-T_5$ as a model prediction for the EHT observations on 2017 Apr 11. The projected angular momentum of the hot spot is indicated with $\mathcal{L}$ and it points into the screen ($i=158^\circ$). Dashed black circles correspond to radii of 12 $r_{\rm g}/D \approx 60\,\mu$as and $3 \sqrt{3} r_{\rm g}/D \approx 26\,\mu$as (critical curve in the Schwarzschild spacetime), where $D$ is the distance toward \sgra. Dashed red lines correspond to the primary and secondary image trajectories on the observer's screen in the case of a point source emission. We report the total intensities and LP of the primary (direct image) and secondary (lensed image) components, as well as those of the unresolved image. Dark magenta ticks indicate the EVPA as it would be observed (corrupted by the Faraday rotation), and green ticks correspond to the model prediction with the Faraday rotation removed. Polarization ticks are shown for the primary and secondary image components, and for a total unresolved source (in the image center). Red and blue curved arrows indicate the receding and approaching side of the hot spot orbit, respectively. }
    \label{fig:comic}
\end{figure*}

The calculations performed for the model of \citet{Vos2022} not only generate full Stokes light curves, but also produce resolved images of the hot spot model. This is illustrated in Fig.~\ref{fig:comic}, for which images of a pure hot spot model, with no accretion disk (no observable shadow component) present, were computed. We used the fiducial model, with a spin of $a_* = 0$, a vertical magnetic field, a Keplerian orbit of $r_{\rm orb} = 11 r_{\rm g}$, and an inclination angle of $i = 158^\circ$. Following Appendix\,\ref{app:orient}, we fixed the PA of the projected hot spot angular momentum $\mathcal{L}$ to 57$^\circ$ east of north. The model of the observed $\mathcal{Q}$-$\mathcal{U}$ loop, shown in the top left panel of Fig.~\ref{fig:comic} (dark magenta continuous line), is rotated with respect to the model of the intrinsic source $\mathcal{Q}$-$\mathcal{U}$ (dashed green line) clockwise because of the Faraday rotation effect. The primary image appears as a polarized blob rotating at $\sim 60\,\mu$as radius with a total intensity $\mathcal{I}_{\rm hsp,0}$ fluctuating between $0.02$ and $0.2$\,Jy. The secondary image appears as a weaker component with a PA shifted by $\sim 140^\circ$, rotating at a radius of $\sim 30\,\mu$as. It is interesting to notice that the Doppler effect on the system appearance is subdominant with respect to the magnetic field and line-of-sight geometry relevant for the synchrotron emission, since the orbital velocity projected onto the line of sight is at most $\sim\!0.1 c$ for this low inclination model. The resolved model is also helpful to highlight the role of the secondary image for a detailed geometry of the small inner $\mathcal{Q}$-$\mathcal{U}$ loop, where the secondary image flux density becomes comparable to that of the suppressed primary image. This effect was discussed by \citet{Gelles2021} in more detail, and it can also be seen in the model comparison shown in the right panel of Fig. \ref{fig:data_model}. Given the relative importance of this effect, we expect that studying detailed geometry of $\mathcal{Q}$-$\mathcal{U}$ loops may, in the future, deliver a proof of the existence of secondary images, thus providing an interesting new test of gravity in the strong-field regime.

If the general appearance of \sgra on 2017 Apr 11 is similar as reported by \citet{SgraP1} on 2017 Apr 6 and 7, that is if it corresponds to a roughly uniform ring of diameter $\sim 52\,\mu$as, then the secondary image location would overlap with the ring of the observable black hole shadow, possibly being swamped by the ring emission. However, the primary image could possibly be detected in the simultaneous VLBI data as a distinct component, a rotating polarized blob appearing between $T_0$ and $T_5$, that is between 9:20 UT and 10:38 UT. Thus, our findings may constitute a prediction or a prior for the EHT VLBI data analysis and interpretation. On the other hand, VLBI data analysis could allow one to measure the size of the hot spot orbit $r_{\rm orb}$, thus constraining the Keplerianity of the flow and breaking the degeneracies from which our analysis suffers.

\section{Total intensity and CP}
\label{app:StokesIV}

The presence of Faraday effects breaks the symmetries of the perfectly optically thin system. In particular, only in the Faraday thin limit (when both Faraday rotation and conversion are weak) does a change of the magnetic field sign $B \rightarrow -B$ result in $\mathcal{V} \rightarrow -\mathcal{V}$ with no effect on $\mathcal{P}$. Since the models matching the expected hot spot polarized flux density $|\mathcal{P}_{\rm hsp}| \lesssim 0.2$\,Jy correspond to a low optical and Faraday depth, the departure from the optically thin model is rather small. For that reason, while we can fix the observed hot spot rotation to the clockwise direction based on the direction of motion on the $\mathcal{Q}$-$\mathcal{U}$ plane, we are not able to robustly differentiate between the orbital plane inclination of $i$ and $180^\circ - i$ based on the $\mathcal{Q}$-$\mathcal{U}$ loop pattern alone. Modeling total intensity and circular polarization could be most helpful in breaking more system degeneracies. 

\begin{figure}[h!]
    \centering
    \includegraphics[width=\linewidth]{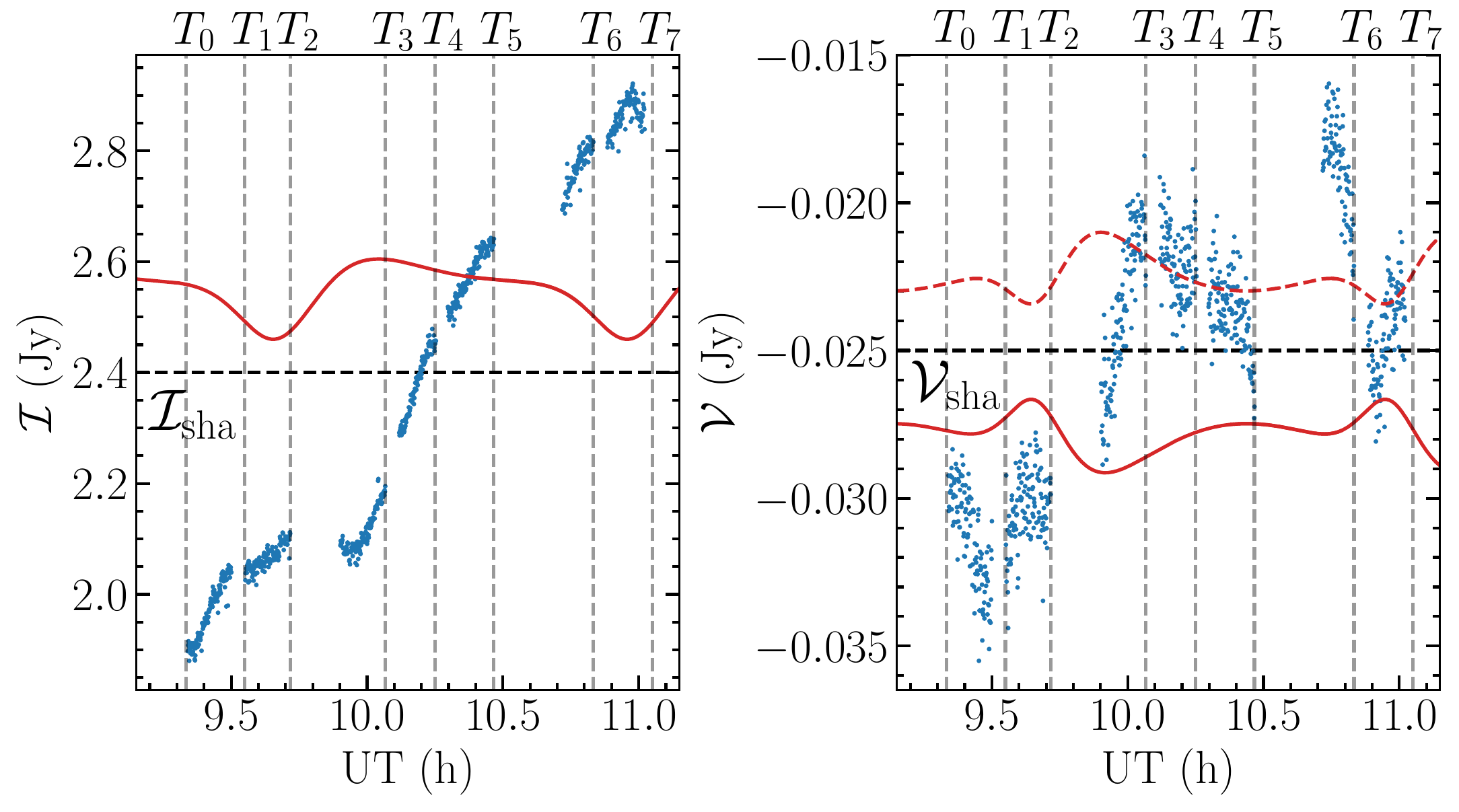}
    \caption{Total intensity $\mathcal{I}$ and circular polarization $\mathcal{V}$ comparison between the fiducial hot spot model presented in Fig.~\ref{fig:data_model} and the observations. A~static source with a total intensity of $\mathcal{I}_{\rm sha} = 2.4$\,Jy and a circular polarization of $\mathcal{V}_{\rm sha} = -0.025$\,Jy, representing the contribution from the observable black hole shadow, was added to the model. For circular polarization, we show the results corresponding to two opposite magnetic field orientations (red continuous and dashed lines).}
    \label{fig:data_model_IV}
\end{figure}

For the fiducial model shown in Fig.~\ref{fig:data_model}, we show the corresponding total intensity and circular polarization light curves compared to ALMA observations in Fig.~\ref{fig:data_model_IV}. A general lack of agreement is not surprising, following the discussion in Section~\ref{sec:general_model}. The total intensity rise during the loopy period may be related to the system cooling down and recovering a normal millimeter emission level, after the flaring event that pushed emission to higher energies \citep{wielgus22}. This flux density increase may affect the hot spot component as well, as we noted in Section~\ref{sec:limitations}. Perhaps the flux density rise is less apparent in LP, as it is countered by the depolarizing effect of the differential shearing.

In the case of CP, there is a hint of correlation between the data and the model prediction for the magnetic field polarity denoted with a dashed line in Fig. \ref{fig:data_model_IV}. We can obtain decent agreement between the data and model by multiplying the model light curves with a factor linearly increasing in time, which would be consistent with a notion of a hot spot cooling down on dynamical timescales, increasing radiative output in millimeter wavelengths. See Appendix \ref{app:delays} for additional hints of plasma cooling being important for the full interpretation of these observations.


\section{Intrinsic source orientation}
\label{app:orient}

\begin{figure}[h!]
    \centering
    \includegraphics[width=\linewidth]{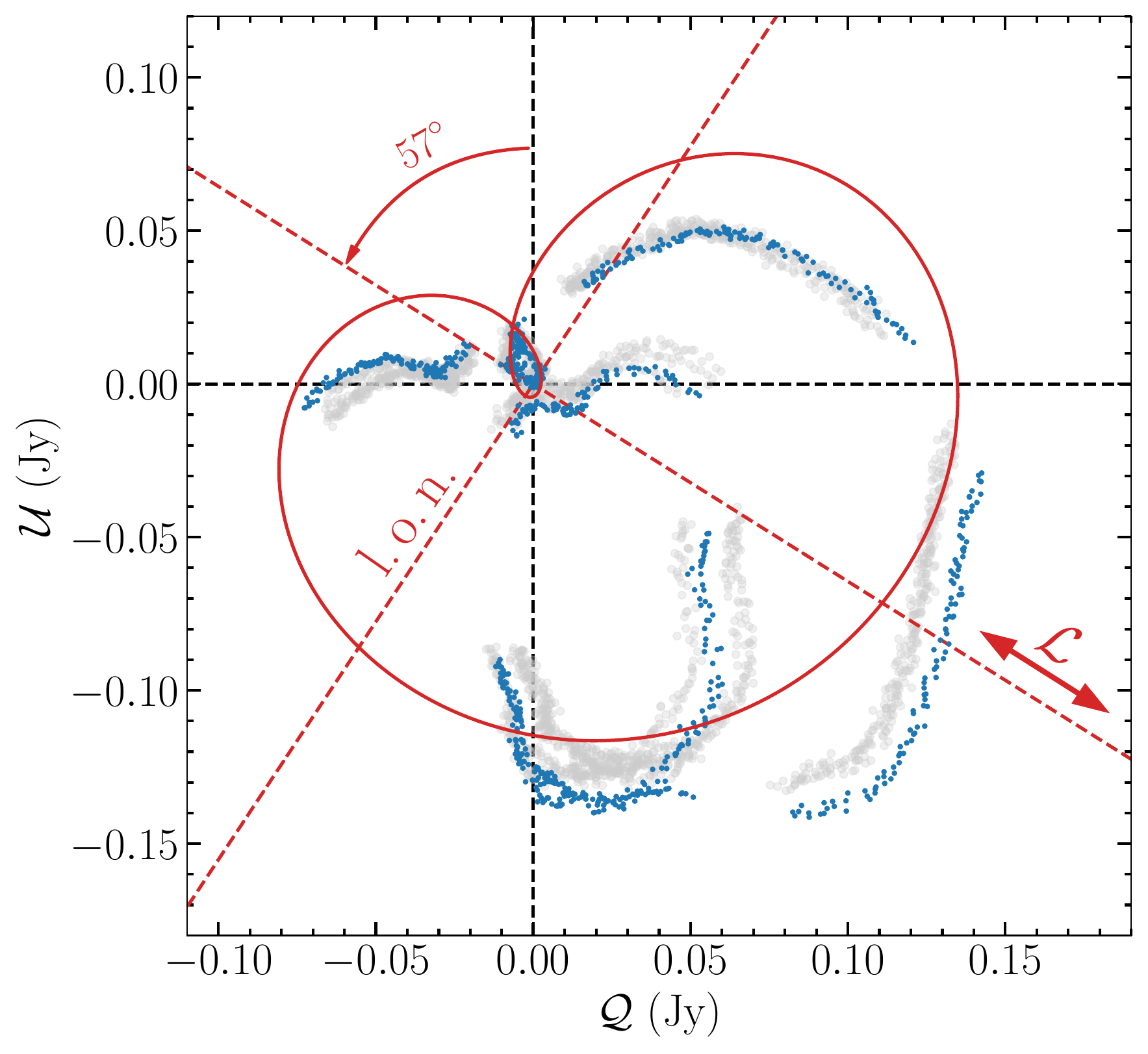}
    \caption{Intrinsic LP in the loopy period, after removing the mean effect of the Faraday rotation. The 229\,GHz band data are shown with a blue color, while the three remaining ALMA frequency bands are shown with a gray color. The implied PA of the line of nodes (l.o.n.) of the hot spot orbit as seen in the sky was estimated at about 147$^\circ$ east of north. The intrinsic PA of the hot spot angular momentum $\mathcal{L}$ was found to be $\sim 57^\circ$ east of north (or 227$^\circ$, given the 180$^\circ$ ambiguity). }
    \label{fig:derotated}
\end{figure}

The observed EVPA of \sgra is corrupted by the time-dependent Faraday rotation, as discussed in Appendix\,\ref{app:RM}. While we argued for a significant intrinsic component of the Faraday screen, potentially varying rapidly both in time and space, the mean value of RM can be used to approximate the contribution of the external Faraday screen. Hence, by correcting the effect of the mean RM, we can obtain the orientation of the $\mathcal{Q}$-$\mathcal{U}$ loop better corresponding to the intrinsic properties of the compact \sgra system. For the mean RM of $-3.28 \times 10^5$\,rad/m$^2$, which we measured on 2017 Apr 11 with ALMA, this corresponds to rotating the observed $\mathcal{Q}$-$\mathcal{U}$ loop pattern counterclockwise by 64.4$^\circ$\ (or rotating the EVPA by 32.2$^\circ$) in the 229\,GHz band. The result of this procedure, applied to the data and to the fiducial model shown in Fig.~\ref{fig:data_model}, is presented in Fig.~\ref{fig:derotated}. In the figure we also rotated the remaining three ALMA frequency bands, and aligned the small loops between bands by removing a frequency-dependent, yet constant in time, component attributed to the LP of the observable black hole shadow $\mathcal{P}_{\rm sha}$. The inter-band consistency between loop patterns, with the small residual differences potentially resulting from the intrinsic Faraday screen effects and dependence of emission on frequency, further confirms the robustness of our results. We infer the PA of the hot spot angular momentum axis projected onto the observer's screen to be $\sim 57^\circ$ east of north, with a 180$^\circ$ ambiguity. These two possible cases correspond to an inclination of $158^\circ$ (as shown in Fig. \ref{fig:comic}) or $202^\circ$ between the line of sight and the hot spot angular momentum vector. Our results are reasonably consistent with the hot spot spin axis PA inferred from GRAVITY IR hot spots data at about 50$^\circ$ \citep[corresponding to their reported line of nodes angle $\Omega$ reduced by 90$^\circ$; ][]{gravity_loops_2018}.


\section{Delay between 229 and 213\,GHz bands}
\label{app:delays}

\begin{figure}[h!]
    \centering
    \includegraphics[width=\columnwidth]{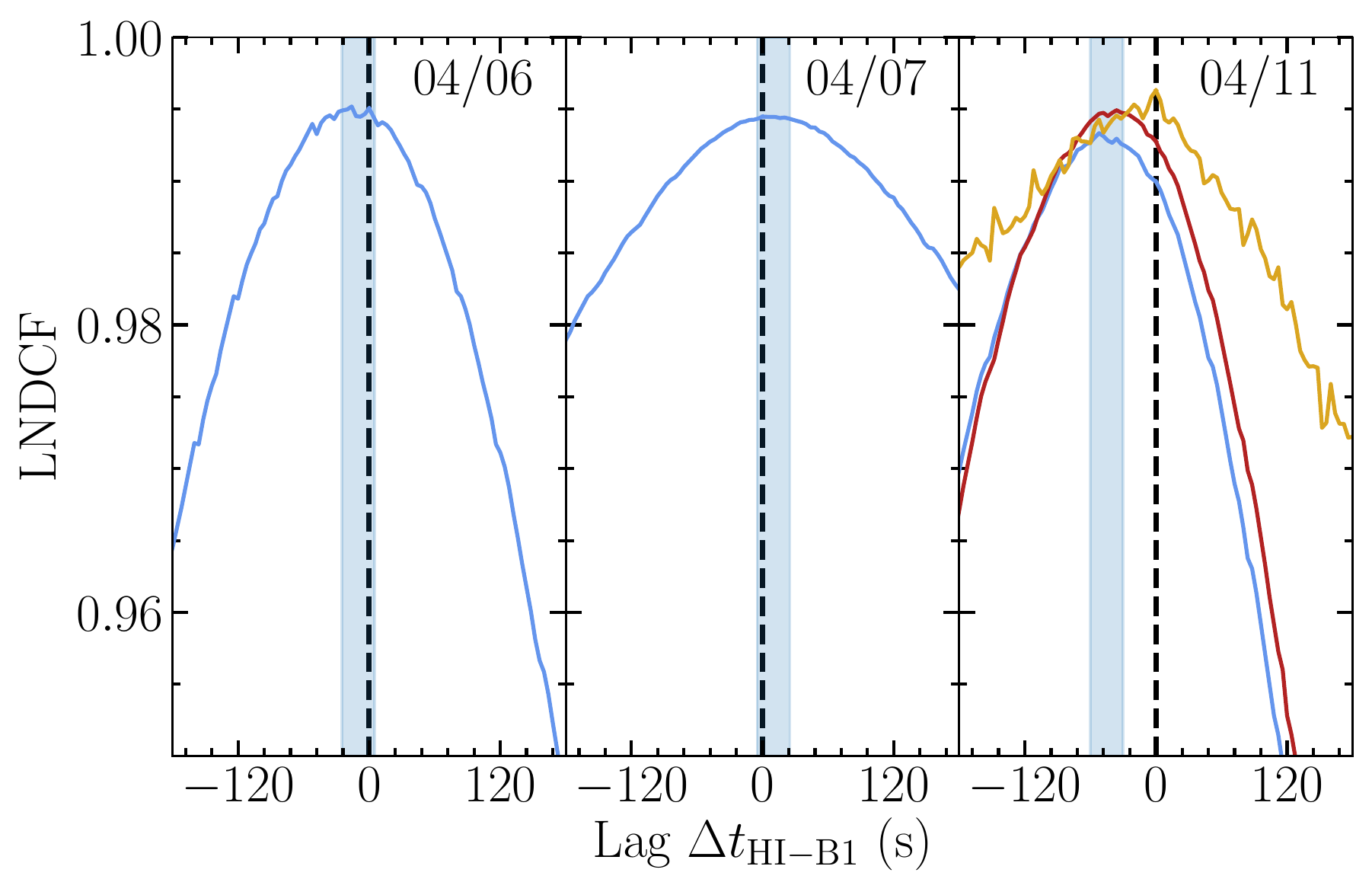}
    \caption{Polarized flux density $|\mathcal{P}|$ correlations between 229\,GHz (HI band) and 213\,GHz (B1 band) frequencies (blue curves). Locations of the correlation maxima with associated uncertainties are indicated by blue shaded regions. For 2017 Apr 6 and 7, no significant delay has been detected. On 2017 Apr 11, the B1 band lags behind the HI band by $45\pm15$\,s. The result is consistent if only the first 2\,h of the Apr 11 observations are used (i.e., before 11.0 UT, corresponding to the loopy period, red curve). The lag disappears if only the latter part of the observations is analyzed (after 11.0 UT, orange curve).}
    \label{fig:correlations}
\end{figure}

We investigated ALMA light curves of \sgra for a presence of delayed correlation between the frequency bands at 229 and 213\,GHz. In the tests presented in the Section 4.5 of \citet{wielgus22} no total intensity delayed correlations for this data set were detected, which was interpreted as an indication of low optical depth of the system. For this study, we repeated the same exercise for the LP component $|\mathcal{P}|$. We used the locally normalized discrete correlation function (LNDCF), as defined by \citet{Lehar1992}. While on 2017 Apr 6 and 7 we found no indication of a delay in $|\mathcal{P}|$, on 2017 Apr 11 we identified a delay of $45\pm15$\,s, with a 213\,GHz signal lagging behind the 229\,GHz. The delay was present during the loopy period on 2017 Apr 11, and it disappeared later that day, see Fig.~\ref{fig:correlations}. While we do not have a reliable quantitative interpretation of this delay, we can speculate that it is a feature related to the hot spot, and not to the accretion flow, given that it only appears in the aftermath of the X-ray flare, and only in LP, dominated by the hot spot component. Since we have excluded the optically thick interpretation corresponding to a difference in the photosphere location at the two frequencies, an alternative source of a delay could correspond to the rapid cooling of the hot spot (which could be radiative and/or adiabatic). The emission shifts toward lower frequencies as the plasma cools down, which can manifest as an apparent correlated delay with a higher frequency signal leading the lower frequency counterpart. More quantitative analysis of the observed delays is beyond the scope of this paper; nevertheless, we notice that interpreting results in a more general modeling framework that incorporates radiative cooling could potentially shed light on the collisionless plasma cooling mechanisms and timescales.

\end{appendix}

\end{document}